\documentclass{aa}  

\usepackage{graphicx}
\usepackage{txfonts}
\usepackage{gensymb}
\usepackage{tikz}

\usepackage{ulem}   
\usepackage{multirow}
\usepackage[dvipsnames,svgnames,x11names]{xcolor}
\usepackage{float}

\usepackage{enumitem}

\usepackage[colorlinks=true,     linkcolor=blue, citecolor=blue, filecolor=blue, urlcolor=blue]{hyperref}

\definecolor{GR}{rgb}{0.8, 0.20, 0.20}

\defcitealias{Cantiello2024}{MC24}
\defcitealias{Tonry2001}{Ton01}
\begin{document}

\title{FAST-SBF: an automatic procedure for the measurement of Surface Brightness Fluctuations for large sky surveys.}
\subtitle{Preliminary test on LSST precursors.}

\titlerunning{FAST-SBF: a new procedure for SBF with LSST}
\authorrunning{Gabriele Riccio et al. }

\author{Gabriele Riccio\inst{1}, Michele Cantiello\inst{1}, Rebecca Habas\inst{1}, Nandini Hazra\inst{1,2,3}, Giuseppe D'Ago\inst{4}, Gabriella Raimondo\inst{1}, John P. Blakeslee\inst{5}, Joseph B. Jensen\inst{6}, Marco Mirabile\inst{1,7,3}, Enzo Brocato\inst{8}, Massimo Brescia\inst{9,10}, Claudia M. Raiteri\inst{11}  }

\institute{INAF-Osservatorio Astronomico d'Abruzzo, Via Mentore Maggini, s.n.c., 64100 Teramo, Italy
\and
National Centre for Nuclear Research, ul. Pasteura 7, 02-093 Warszawa, Poland
\and 
Gran Sasso Science Institute, Viale Francesco Crispi 7, 67100 L’Aquila, Italy
\and
Institute of Astronomy, University of Cambridge, Madingley Road, Cambridge, CB3 0HA, UK
\and
NSF’s NOIRLab, 950 N. Cherry Avenue, Tucson, AZ 85719, USA
\and
Utah Valley University, Orem, Utah 84058, USA
\and
European Southern Observatory, Karl-Schwarzschild-Strasse 2, 85748 Garching bei München, Germany
\and  INAF-Osservatorio Astronomico di Roma - Via Frascati 33, 00040, Monte Porzio Catone
\and
Department of Physics "E. Pancini", University Federico II, Via
Cinthia 6, 80126, Napoli, Italy
\and
INAF-Osservatorio Astronomico di Capodimonte, Via Moiariello
16, 80131 Napoli, Italy
\and
INAF - Osservatorio Astrofisico di Torino, via Osservatorio 20,
10025 Pino Torinese (TO), Italy
}

  \abstract
   {The Surface Brightness Fluctuation (SBF) method is one of the most reliable and efficient ways of measuring distances to galaxies within 100 Mpc, a matter of crucial importance for all fields of astrophysics. While recent implementations have increasingly relied on space-based observations, SBF remains effective when applied to ground-based data. In particular, deep, wide-field imaging surveys with sub-arcsecond seeing conditions allows us for accurate SBF measurements across large samples of galaxies. With the upcoming next generation wide-area imaging surveys, the thousands of galaxies suitable for SBF measurements will give us the opportunity to constrain the 3D structure of the local universe. }
   {We present FAST-SBF, a new Python-based pipeline for measuring SBF, developed to support the analysis of large datasets from upcoming wide-field imaging surveys such as LSST, Euclid, and Roman. The procedure, still in the testing and development stage, is designed for automation and minimal user intervention, offering a fast and flexible approach to SBF distance estimation. }
   {We validate the performance of the procedure on high-quality imaging data from the Hyper Suprime-Cam Subaru Strategic Program (HSC-SSP), a precursor to LSST, analyzing a sample of both luminous early-type galaxies and fainter dwarfs. Our measurements are also compared with recent results from the Next Generation Virgo Cluster Survey (NGVS) and with the SPoT stellar population synthesis models.}
   {The results show excellent agreement with published distances, with the capability of measuring the SBF signal also for faint dwarf galaxies. The pipeline allows the user to completely analyze a galaxy in relatively short time ($\approx$ minutes) and significantly reduces the need for user intervention.
   The FAST-SBF tool is planned for public release to support the community in using SBF as a distance indicator in next-generation surveys.}
    {}

\maketitle

\section{Introduction}
\label{sec:Introduction}

An accurate determination of distance is essential for deriving almost all the key properties of cosmic structures and objects, such as mass, luminosity, and size. Furthermore, resolving the three-dimensional structure of galaxies within clusters requires distance measurements for individual targets with a precision better than the ratio between the depth of the cluster and its mean distance, typically corresponding to uncertainties of less than $\sim$0.1 mag in distance modulus (5\% in distance) in the local universe \citep{Mei2007, Blakeslee2009, Cantiello2018, Rodriguez2024}. Only a few distance indicators can achieve such precision. Among them, one which offers both the broad applicability necessary to resolve the depth of galaxy clusters out to the Hubble flow and the capability to be applied to a significant fraction of galaxies within the same environment is the surface brightness fluctuation method (SBF). For a detailed and updated description of the method, we refer to the review from \citet{Cantielloreview}. 

Since the first definition of the SBF technique by \cite{Tonry1988}, a major observational effort has been made to improve the accuracy and precision of the method both as a distance indicator \citep[e.g.][]{Tonry1997,Tonry2001,Cantiello2007,Mei2007,Blakeslee2009,Cantiello2018,Jensen2021,Cantiello2024} and as a diagnostic for the properties of the unresolved stellar populations in galaxies \citep[e.g.][]{Buzzoni1993,Worthey1994,Cantiello2003,Raimondo2005}.
The precision of the method depends on the quality of the measurements (image quality, background, extinction, and photometric uncertainty) and the wavelength-dependent scatter that arises from variations in the stellar populations of the galaxies themselves. Overall, in the optical passbands, the errors of the measurements can achieve statistical uncertainties on the SBF magnitude of $\lesssim 0.1$ mag for ground-based observations \citep[e.g.][]{Cantiello2024} and $\lesssim 0.05$ mag for space-based observations \citep[e.g.][]{Blakeslee2009}. SBF in the near-IR has some advantage over optical bands, as the amplitude of the fluctuations is stronger, the impact from undetected sources is less dramatic than in optical bands, and the effect of residual dust is negligible in most cases. Recent studies based on HSC/WFC3 data achieved SBF measurement precisions of $\sim 0.08$ mag in the best cases \citep{Jensen2021}, not including the systematic uncertainty in the zero point calibration.

These characteristics make SBF a promising distance indicator for the  forthcoming generation of wide-area imaging surveys such as Rubin, Euclid, and Roman, which will image many thousands of galaxies suitable for SBF measurement. In particular, the \textit{Legacy Survey of Space and Time} \citep[LSST;][]{Ivezic2019} of the Vera C. Rubin Observatory will scan about 18\,000 deg$^2$ of the southern sky reaching an $i$-band depth of $\approx26.5$ for the ten-year coadded images with a median FWHM $\sim 0\farcs6$, which will allow us, in principle, to estimate distances with SBF up to $\sim\,$50 Mpc, possibly farther for selected objects \citep{Cantielloreview}. This represents an important opportunity to constrain the 3D structure of the local universe. Considering the large amount of data that will be produced by Rubin and other deep surveys, it is essential for the SBF analysis to be carried out in a fast, robust, and automatic way. However, the existing SBF measurement procedures require extensive user intervention.

In this work, we present the first results from a new pipeline for the measurement of the SBF signal, specifically developed to meet the requirements of speed and automation. We test the procedure on the Hyper Suprime-Cam Subaru Strategic Program (HSC-SSP) survey data, a precursor to LSST, and with the recent measurements from the Next Generation Virgo Cluster Survey \citep[NGVS][]{Ferrarese2012,Cantiello2024}. The paper is organised as follows. In Section~\ref{Sec:Data} we describe the data and the HSC-SSP survey. In Sections~\ref{Sec:Methodology} and~\ref{sec:Results} we present the SBF measurement procedure and then compare our measurements to previous results and to predictions from stellar population models. Our conclusions are presented in Section~\ref{Sec:Conclusions}. 

\section{Data and sample selection}
\label{Sec:Data}

This study is mainly based on imaging data from the second data release (PDR2) of the HSC-SSP. The primary motivation for using HSC data lies in its intrinsic similarities with LSST. Both HSC-SSP and LSST are designed to conduct deep, wide-field optical surveys using similar photometric bands ($grizy$ for HSC and $ugrizy$ for LSST) and, most importantly, the data reduction pipelines developed for LSST are built upon the same framework as the HSC science pipelines \citep{Ivezic2019}. The accuracy of the SBF signal also depends on the data reduction strategies adopted \citep{Cantiello2005,Mei2005b}. Thus, HSC-SSP data serve as an exceptionally valuable resource for the optimization and the automation of the SBF measurements procedure for the LSST dataset. 

Full details on the HSC-SSP survey strategy and image processing are provided by \cite{Aihara2019}. We briefly summarize the specifics of the survey. The Subaru Telescope is a wide-field optical-infrared imaging telescope with a 8.2 meter aperture, located at the Mauna Kea Observatory in Hawaii. The telescope is equipped with the 870 megapixel Hyper Suprime-Cam \citep{Miyazaki2018}, which covers a 1.5 deg$^2$ field-of-view with a pixel scale of $0\farcs17$. The HSC-SSP multi-band imaging survey was designed to map a fraction of the sky in three layers: Wide (1400 deg$^2$), Deep (27 deg$^2$) and Ultradeep (3.5 deg$^2$). For this work, we choose the PDR2 Wide layer, which covers $\sim$1100 deg$^2$, reaching a $5\sigma$ limiting magnitude for the coadded images of 26.6, 26.2, 26.2, 25.3 and 24.5 mag in the \textit{g}, \textit{r}, \textit{i}, \textit{z} and \textit{y} bands respectively, with a median \textit{i} band FWHM of $0\farcs58$. We use the stacked images from the PDR2 instead of the more recent and spatially complete PDR3 because the background subtraction strategy adopted for the PDR3 images can severely oversubtract the galaxy halo for bright and extended galaxies \citep{Aiharapdr3}, potentially biasing the SBF measurement, which relies on an accurate model of the galaxy light distribution. The images obtained from the HSC-SSP archive are calibrated and normalized with identical zero-point in all bands ($m_{\mathrm{zp}}= 27$). The magnitude of the sources is derived as $m=-2.5\log_{10}(f)+m_{\mathrm{zp}}$, where \textit{f} is the flux measured on the image.

\begin{figure*}
    \centering
    \includegraphics[width=1\hsize]{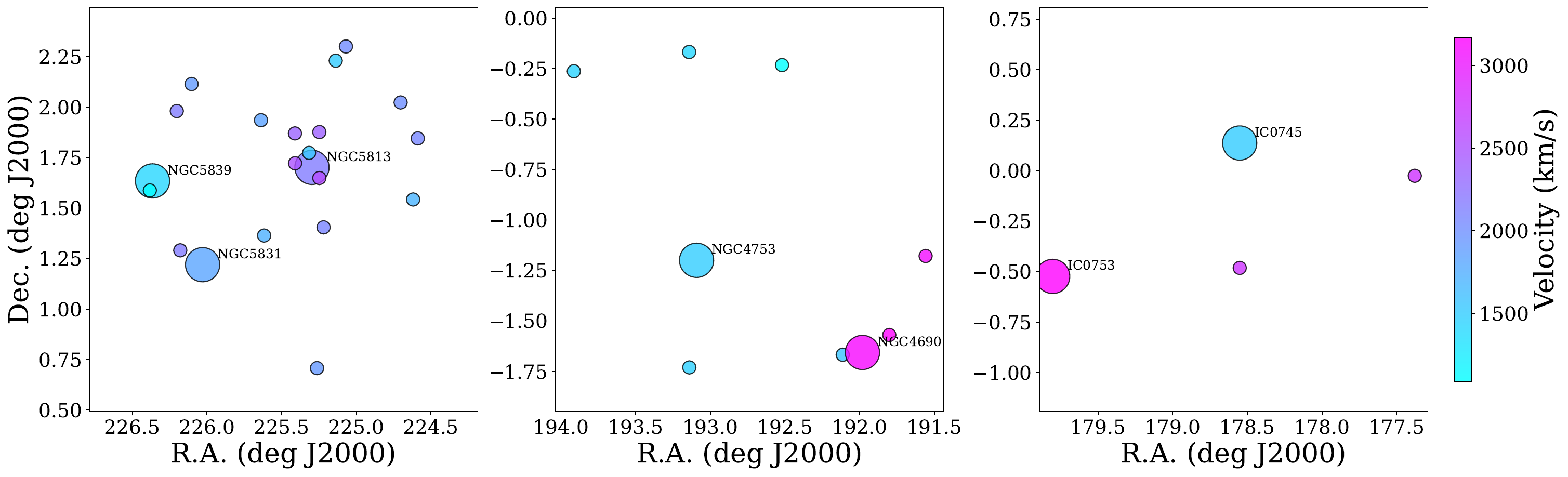}
    \caption{ Spatial distribution of the galaxies in the three fields we studied. The color represents their velocity in respect to the CMB frame. Small circles represent dwarf galaxies, while bright galaxies are represented with bigger size and are labeled. The box size is $1.3\times1$ deg and is the same for all the panels.}
    \label{fig:spatial_distr}
\end{figure*}

From the PDR2 sample, we identified a subset of galaxies with existing SBF measurements in the literature, focusing on large galaxy catalogs
\citep{Tonry2001,Blakeslee2009,Jensen2015}. At the time the cross-match of the PDR2 archive with such catalogs was performed, we found five matches for galaxies observed both in the $i-$ and $g-$band, all of which are from \citet[][hereafter \citetalias{Tonry2001}]{Tonry2001}. We chose the $i-$ and $g-$bands for coherence with the existing literature, using the $i$-band for SBF measurements, and the combined bands to get the $g-i$ color required for estimating the absolute SBF magnitude (see Sect. \ref{3.6}). \citetalias{Tonry2001} presented SBF distances to $\sim$300 galaxies and 20 nearby groups. The five selected galaxies are distributed across three distinct sky regions, shown in Fig. \ref{fig:spatial_distr}. 
We further expanded our sample by visually inspecting the $\approx 1.5\times1.5$ deg area around the galaxies to identify additional candidates potentially associated with the reference galaxies and suitable for SBF analysis. In total, we identified 31 sources, two of which are relatively bright early-type galaxies while the remaining are fainter dwarf-like galaxies, all with no previous distance available from the literature. The majority of the dwarf galaxies identified are also observed in the MATLAS survey \citep{Duc2015, Habas2020,Poulain2021}. The expanded sample offers the opportunity to assess the performance of our SBF measurement code on fainter objects.
While we did not impose strict morphological selection criteria, we prioritized galaxies with smooth morphology. For this reason, our sample is dominated by galaxies with early-type morphologies. The remaining systems are later types, but usually with the presence of a prominent bulge or a region showing a smooth profile. This regularity is key to observe an effective SBF signal for distance estimates, as early-type galaxies with smooth profiles minimize artificial residuals arising from the modeling procedure (see Section \ref{sec:3.2}) and limit the contribution of young stellar populations that could bias the SBF calibration.  \citep[e.g.][]{Moresco2022}.
Table\,\ref{table:1} presents the properties of the galaxies included in our sample.

In addition to the comparison with the \citetalias{Tonry2001} measures, we also run our new pipeline on data with more recent ground-based SBF measurements. As demonstrated in \citet{Cantiello2018}, the roughly thirty-year-old SBF survey by Tonry \& collaborators contains some outliers, especially for the most distant targets in their sample. For this comparison, we included additional data from the Next Generation Virgo Cluster Survey \citep[NGVS,][]{Ferrarese2012}, which were previously studied in \citet{Cantiello2024}.
The NGVS survey obtained $ugriz$ band imaging data of 104 sq. degrees in the Virgo cluster using CFHT MegaCam data, with $i$-band observations optimized for the SBF measurements. Here, we run our pipeline on a selection of 20 galaxies from the NGVS with different total magnitudes and colors, to verify the results of our procedure. The data are the same used in \citet{Cantiello2024}, and were obtained from the NGVS archives, now public; the images are normalized and calibrated such that the zero-point is $m_{\mathrm{zp}}=30$ mag in all bands.

\section{SBF measurements with FAST-SBF}
\label{Sec:Methodology}

\begin{figure}
    \centering
    \includegraphics[width=0.97\hsize]{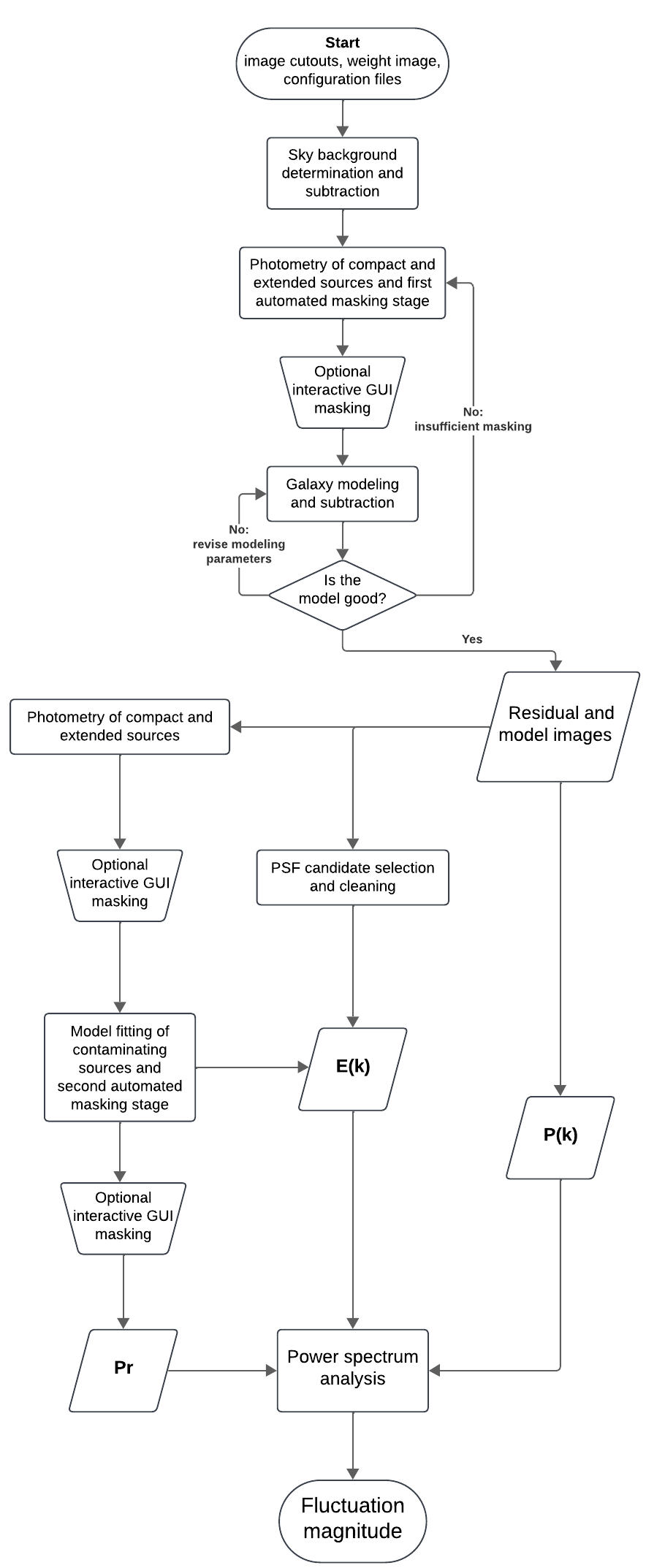}
    \caption{Flow chart of the main process of the FAST-SBF pipeline.}
    \label{fig:flow_chart}
\end{figure}

To measure the SBF magnitudes, we follow an approach similar to the well established procedures already developed in previous works \citep[e.g.][]{Blakeslee2009,Blakeslee2021,Cantiello2018,Cantiello2024,Jensen2021,Mei2005b}, automated as much as possible to efficiently handle large datasets. We have made an effort to use publicly available Python packages for this purpose.
Below, we provide an overview of the technique, adding some practical notes, and detail the changes and advancements introduced. Figure \ref{fig:flow_chart} shows the flow chart of the adopted procedure. 
One of the main purposes of the present work is to show the progress in the automation and Python porting of the code, named {\sc{Flexible Automated Self-contained Tool for SBF}} (FAST-SBF), and establish its consistency with previous literature results. The pipeline is currently under active development, and certain components will be modified and improved prior to its public release
\footnote{We encourage interested users to contact the authors for access and further information.}. 
The procedure is developed to work for any optical to near-IR passband. The band-dependent parameters (e.g. GCs and background galaxies luminosity function fitting, extinction correction, instrumental zero-point, etc.) are user defined, and are provided in a configuration file. 

\subsection{Sky background}
\label{sec:bkg}
The first step of our procedure entails accurately determining the sky background level to be subtracted from the image of the galaxy cutout\footnote{Cutout sizes were chosen to include sufficient empty area for background determination. These are obtained using the HCS-SSP image cutout service (\url{https://hsc-release.mtk.nao.ac.jp/das_cutout/pdr3/}) and have typical sizes of 20$\times$20 arcmin for bright galaxies, with smaller cutouts for fainter ones.}. Any uncertainty in the sky level directly impacts the SBF signal and the galaxy color measurements, as well as the associated errors. Typically, the sky uncertainty contributes $\leq0.02$ mag of the total error on the SBF, corresponding to less than 1\% on the distance \citep{Cantielloreview}.
The survey data used here have already been sky-subtracted in the image-processing stage. However, a residual sky value might be present and needs to be estimated and subtracted. 

\begin{figure}
    \centering
    \includegraphics[width=1\hsize]{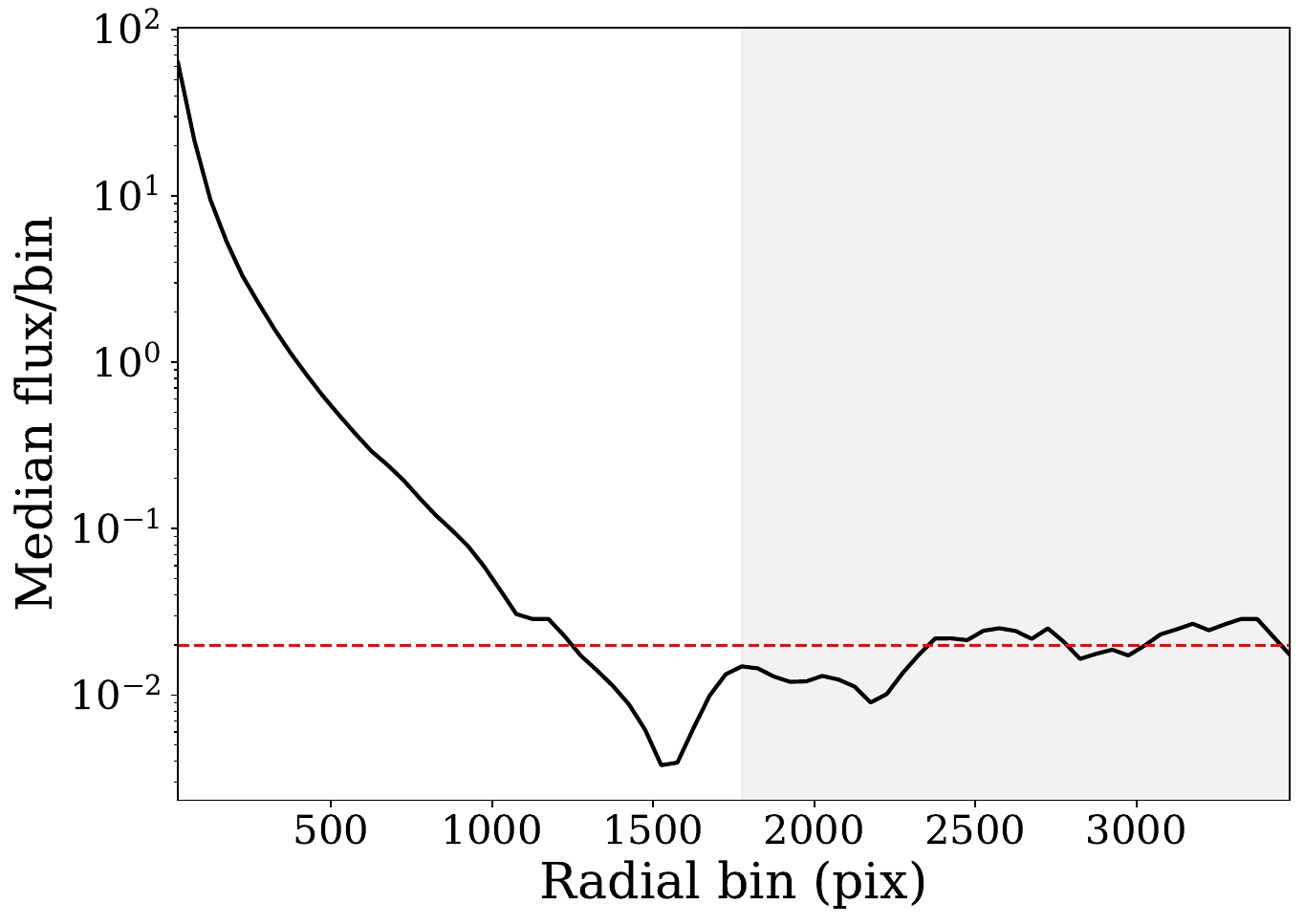}
    \caption{Example of the procedure to estimate the $i$-band background of NGC\,5831. The black solid line shows the radial profile of the flux in the input image, azimuthally averaged in circular bins of 50 pixels. The shaded gray area is the region where the background is estimated. The red dashed line represents the background estimated with the method described in the text. }
    \label{fig:bkg}
\end{figure}

The tests and results presented here rely on wide-field imaging data, therefore the sky can generally be determined far from the galaxy center. 
We first estimate the median flux in circular annuli concentric to the galaxy, extending outward. 
The sky background value is then determined as the sigma-clipped median value of the radial flux profile, evaluated in a region away from the galaxy core, where the radial flux profile flattens out. The uncertainty on the background, $\sigma_{\mathrm{sky}}$, is derived from the median absolute deviation (MAD) estimated within the same region (for a normal distribution, $\sigma=1.483\times MAD$, where $\sigma$ is the standard deviation). Figure \ref{fig:bkg} shows an example of the approach adopted for NGC\,5831, one of the five galaxies with HSC-SSP data and in the \citetalias{Tonry2001} sample.
This method may be less reliable when the galaxy covers most of the field. The present version of the code is tailored for the wide-field surveys like the HSC-SSP and NGVS, hence estimating the sky does not generally represent a problem. 
Future versions of the code, already in testing stage, will address the issues arising from observations where the galaxy covers entirely the detector (like massive elliptical galaxies in the local universe imaged with the Hubble or James Webb space telescopes).

\subsection{Galaxy modelling and subtraction}
\label{sec:3.2}
The SBF signal is defined as the ratio between the intrinsic variance of the stellar light distribution and the mean surface brightness within a given region of the galaxy. Thus, a smooth isophotal model of the galaxy surface brightness after sky subtraction is needed; the resulting model frame corresponds to the first moment of the light distribution. In order to perform the modeling of the galaxy, we first derive the photometry of compact and extended sources using SExtractor \citep{Bertin1996}. The photometric catalog is used to mask the brightest sources in the frame, adopting the appropriate source shape\footnote{Circular shape for compact sources and elliptical for the extended ones based on SExtractor's geometric parameters (e.g \textit{$A\_WORLD$}, \textit{$B\_WORLD$} and \textit{$THETA\_WORLD$)}.}, magnitude cuts (in the same magnitude scale of the images), and masking radius scaled to source magnitude. 

The background subtracted masked frame is then used to derive a 2D model of the galaxy surface brightness, with the ELLIPSE routine \citep{Jedrzejewski1987} from the Photutils\footnote{\url{https://photutils.readthedocs.io/en/stable/index.html}} package. This routine offers reliable results with minimal user intervention, but it is
computationally time consuming for extended galaxies ($>10^5$ pixels). Moreover, a relevant limitation occurs when modelling isophotes heavily contaminated by globular clusters, bright companions, background sources, dust patches, or bright stars. Most of these sources are automatically masked using the compact and extended source catalogs.
However, some objects (like extended dust patches) or features (e.g. bright stellar halos or diffraction spikes) may evade the preliminary masking stage. To address this, we have incorporated an optional manual masking process, based on an interactive graphical interface.

Usually, the model-subtracted image shows residual patterns from the model subtraction stage, especially in cases where
central dust lanes, shells, bars or spiral sub-structures are present in the target galaxy. 
As will be discussed later (Sect. \ref{sec:fluctuations}), the presence of large-scale residual patterns is not a limitation for the procedure.
Nevertheless, the large-scale residuals still present in the frame after subtracting the galaxy model are removed using a background map obtained with SExtractor\footnote{For the data presented here, we adopt a background map derived running sextractor adopting mesh size $BACK\_SIZE=10 \times FWHM$ and $BACK\_FILTERSIZE=3$ \citep[e.g.][]{Jordan2009}.}.
In contrast, small-scale residual structures, such as localized rings, shells, or dust clumps, can introduce biases in the measurement and are therefore masked during the analysis (see Sect. \ref{sec:fluctuations}).  
Hereafter, we refer to the image with the sky, galaxy model, and large-scale residuals subtracted as the residual frame. Top panels of Fig. \ref{fig:power_spectrum} show the procedure described in this section for the case of NGC\,5813. The images show the $i$-band  galaxy cutout, the residual, and residual masked image after the automatic masking of the detected sources (left to right). The right panel also indicates the annulus used for the SBF measurement.

\subsection{Point-Spread Function selection}
\label{sec:3.3}
In the ideal case, without atmospheric or instrumental blurring and in the absence of external sources of fluctuation beyond the stellar counts, the SBF signal would appear indistinguishable from the instrumental white noise. In practice, the convolution with the point-spread function (PSF) introduces correlations between adjacent pixels. The SBF signal is measured in Fourier space, where the power spectra of the signals convolved in the physical space (PSF and stellar variance) are multiplied by each other (see Sect. \ref{sec:fluctuations}). For this reason, a key ingredient for accurate SBF measurements is the image PSF \citep[see][]{Blakeslee1999, Moresco2022,Cantielloreview}. Owing to the wide-field images available, in this work we characterize the PSF relying on local stars.

We compile a preliminary library of PSF reference stars by extracting bright and isolated point sources from the regions around the target galaxies obtained by running SExtractor on the residual frame. For each field, we select compact objects with magnitudes ranging from half a magnitude fainter than the saturation limit ($m_\mathrm{i}\sim$18 mag for all the fields considered), down to two magnitudes fainter.At the distance of our sample and the given image resolution, globular clusters (GCs) appear as point sources and are therefore indistinguishable from stars. However, the magnitude range of the GCs host in our sample galaxies (see Sect. \ref{sec:3.4}) is significantly fainter than the range adopted for the PSF selection. Therefore, when constructing the PSF sample, we do not expect any confusion between GCs and stars. The magnitude range is defined in the parameter configuration file, hence it can be extended for fields lacking isolate, bright and compact PSF candidates. 
To minimize contamination from the galaxy light, we select stars located at a distance of $r_{\mathrm{PSF}}=10 R_\mathrm{e}$ effective radii from the target. To identify isolated stars only sources at least 10 FWHM from any other detected source are kept. The middle row in Fig. \ref{fig:power_spectrum} shows an example of the PSF selection process around NGC\,5813. Some of the selected stars exhibit a residual contamination from undetected faint sources or residual image artifacts. To refine the selection we adopted a two-steps strategy. First we perform an azimuthal average of the normalized flux for each candidate,
then the profiles are combined using a sigma-clipping algorithm to build a reference PSF (red solid line in Fig. \ref{fig:power_spectrum}, middle right panel). 
The selection of best PSF candidates is made by comparing the azimuthal profile of each individual star to the reference PSF, through a chi-square minimization.
Second, we further clean the PSF cutout by replacing all outlier pixels beyond a given distance from the PSF core (20$\times$FWHM for the data analyzed here) with a pixel value extracted randomly from the distribution along the same radial bin. Finally, the background is estimated around each individual PSF, and is subtracted from the PSF cutout.

In this work we select an average of 10 good PSFs for each galaxy. A sub-sample of the good PSFs identified for the case of NGC\,5813 is shown in Fig. \ref{fig:power_spectrum}, highlighted with red borders (middle, left panel). The PSF selection and cleaning procedure can be tailored by changing the isolation criteria, the magnitude ranges, SExtractor stellarity index, etc. The entire procedure of PSF selection is completely automatic and is computationally quite efficient ($\approx$10 seconds to run on each galaxy in our reference sample). To obtain the amplitude of the fluctuation signal, the power spectrum analysis is performed independently as many times as the number of good PSF identified (see Sect. \ref{sec:fluctuations}).

\subsection{Residual power from globular clusters and background galaxies}
\label{sec:3.4}

In addition to the stellar counts fluctuations that generate the SBF signal, the total fluctuation signal is also affected by contamination from compact and extended sources, mainly GCs and background galaxies \citep[e.g.][]{Blakeslee1995,Tonry1997}. 

The first step to reduce the contamination from the non-stellar fluctuations is to mask all detected GCs and background sources. We use the photometric catalog derived running SExtractor on the residual frame (Sect \ref{sec:3.3}) to mask all such contaminating sources. Sources fainter than the detection limit may still represent a source of contamination to the total fluctuations measured. The procedure we adopted with FAST-SBF to correct for the contamination from undetected sources is a python version of the well-established strategy adopted in previous works \citep[e.g.]{Blakeslee1995, Cantiello2005, Jensen2015}. This step entails modeling the GC luminosity function (GCLF) as a Gaussian distribution and the background galaxies luminosity function as a power law, using the photometric catalog derived in the previous step. The parameters adopted for fitting both components to the luminosity function are user-defined. In particular, we assumed a Gaussian width in the range $\sigma_{\mathrm{GCLF}}=0.9-1.4$ mag for the dwarf/bright galaxies \citep{Villegas2010} and an $i$-band GCLF peak magnitude of $M_\mathrm{i}=-8.0$ mag \citep{Harris2002, Durrell2014}. The $i$-band background galaxy luminosity function is fitted assuming a power law with slope of $\gamma=0.34$ \citep{Bernstein2002,Cantiello2005}. 

The residual power contribution, $P_\mathrm{r}$, arising from faint undetected GCs and background galaxies, is then computed by integrating the composite LF beyond the photometric completeness limit \citep{Blakeslee1995,Cantiello2007}. This residual power was subtracted from the total fluctuation power, $P_0$ (Sect. \ref{sec:fluctuations}), to isolate the stellar SBF signal: $P_\mathrm{f}=P_0-P_\mathrm{r}$. 

Previous studies have shown that in optical passbands the GCLF can be reliably constrained with observations reaching $>$0.5 mag fainter than the peak of the GCLF \citep{Cantielloreview}. At such depth, the residual contribution from unmasked, contaminant sources to the SBF signal is reduced to $<$10\%, and it becomes negligible if the completeness limit extends >1 mag beyond the GCLF peak. For the bright galaxies in this study, we have an average contamination of $P_\mathrm{r}/P_0\sim$ 16\%. GC contamination does not represent an issue for the dwarf galaxies inspected here, because their GC populations are typically negligible \citep[e.g.,][and references therein]{Marleau2024}. Thus, for the dwarfs,  we neglected the contribution from GCs to $P_\mathrm{r}$, and only accounted for the contribution from  background galaxies in the correction of the SBF signal.  

A common ingredient in SBF measurements pipelines is the modified weight map used for the (second, in our case) SExtractor run. This is required to account for the high-amplitude fluctuations close to the central regions of the galaxy where the high surface brightness causes the SExtractor to detect the SBF star-count fluctuations as if they were GCs, which can then end up masked and bias $P_\mathrm{r}$. To prevent false detections of the SBF signal as spurious sources, especially near the core of the brightest galaxies, the variance map used by SExtractor includes the galaxy model profile scaled by a constant factor, $\kappa$, to balance noise suppression and source detection \citep[see e.g][]{Jordan2004}. Through empirical calibration, we determined $\kappa\sim 0.02$  for the NGVS sample and $\kappa\sim 0.002$ for the HSC-SSP dataset. Future developments of FAST-SBF will include a procedure for determining $\kappa$.

The process to mask sources and estimate $P_\mathrm{r}$ is fully automated beyond some iteration required to determine $\kappa$. This step is quite efficient in terms of computational time and for the convergence of the fitting algorithms. The parameter settings in the detection algorithm and luminosity function fitting are strictly telescope- and band-dependent, and thus are flexible and user defined.

\subsection{Fluctuation magnitudes}
\label{sec:fluctuations}
The previous steps produced the residual masked frame, a list of reference PSF stars and a table containing the input for estimating the residual fluctuation due to undetected sources, $P_\mathrm{r}$. The next step is to identify the region of the galaxy to be used for the measurement of the SBF signal. 

For bright galaxies, the fluctuation signal is strongest in the central regions, but the fraction of useful pixels in this region can be relatively small (because of 
poor modeling of the steep brightness profile gradients, the presence of contaminating dust, core saturation,  etc.) and the $P_\mathrm{r}$ correction is largest due to the central concentration of GCs. In the outer galaxy regions, and for fainter galaxies, the S/N ratio of the fluctuation signal is typically lower, making the measurement more vulnerable to background variations and increased noise levels. However, the region available is, typically, wider and the contamination from GCs and background galaxies is lower. There is therefore a region between the two extremes where the SBF signal is optimally measured. 

In our procedure, the SBF signal is measured within a single annulus, individually defined for each galaxy. At present, the inner and outer radii of the measurement annulus are still not well automated, although they are relatively stable for galaxies with similar magnitudes and morphologies.
The outer boundary of the annulus, $r_{\mathrm{out}}$, is chosen such that the ratio between the $rms$ of the galaxy flux and the sky+detector background $rms$ is $S/N(r_{\mathrm{out}})\geq5$, for bright galaxies\footnote{For some dwarf galaxies we relax this constraint, reaching $S/N(r_{\mathrm{out}})\geq1$. As will be discussed in Sect. \ref{sec:Results}, these objects are classified as poor-quality measurements, and kept in the sample just to confirm group membership.}. The inner boundary, $r_{\mathrm{in}}$, is set to exclude regions potentially affected by contamination, such as dust lanes or residual modeling artifacts. Additionally, the optional interactive graphic interface is run again at this stage to manually mask any remaining contamination left out in the previous steps. 

The Fourier spatial power spectrum of the annular residual-masked frame is then derived using the \textit{Numpy} \citep{harris2020array} package Fast Fourier Transform\footnote{\tiny \url{https://numpy.org/doc/stable/reference/generated/numpy.fft.fft2}}  and the resulting 2D image is azimuthally averaged to obtain the profile $P(k)$ as a function of the wavenumber $k$.
The fluctuation amplitude, $P_0$, is estimated by fitting the equation:
\begin{equation}
    P(k)=P_0 \times E(k) + P_1.
    \label{eq:power}
\end{equation}

where $E(k)$ is the (azimuthally averaged) power spectrum of the PSF (normalized to have a total energy of one) convolved with the mask \citep{Tonry1990}, 
and $P_1$  is the instrumental white noise component arising from photon shot- and readout-noise, independent of the wavenumber $k$. The convolution is performed using the \textit{Numpy} package \textit{convolve}\footnote{\tiny \url{https://numpy.org/doc/stable/reference/generated/numpy.convolve}}, while for the fit of $P(k)$ to $E(k)$ we use the linear regression method with the \textit{Scipy} \citep{2020SciPy-NMeth} package \textit{curve\_fit}\footnote{\tiny \url{https://docs.scipy.org/doc/scipy/reference/generated/scipy.optimize.curve_fit.html}}. The parameters $P_0$ and $P_1$ are then obtained as outputs of this fitting procedure. The power $P_0$ is converted to apparent SBF magnitude $\overline{m}$ using the relation:
\begin{equation}
    \overline{m}=-2.5 \log_{10} (P_0-P_\mathrm{r}) + m_{\mathrm{zp}} 
\end{equation}
where $m_{\mathrm{zp}}$ is the photometric zero point for the image. Finally, the band-dependent foreground extinction correction is applied, using $A_\mathrm{i}=1.684 \times E(B{-}V)$ \citep{Schlafly2011}. 

The fitting procedure for Eq.\,\ref{eq:power} can be replaced by a robust fitting algorithm based on the \textit{random sample consensus} method \citep[RANSAC,][]{RANSAC}. In normal cases, the RANSAC routine produces results equivalent to the default fitting method, and is more time consuming than the default method. However, this option is useful in cases where extreme outliers are present in the $P(k)$ or $E(k)$ radial profiles.

The power spectrum $P(k)$ for the residual image can be distorted by large-scale background patterns from the galaxy model fitting procedure, which often adds power at low wavenumbers. Similarly, the flat noise power spectrum ($P_1$) can be tilted when the pixel interpolation kernel used in combining images produces correlations between pixels. We therefore exclude both low and high wavenumbers when fitting for $P_0$ Eq.,\ref{eq:power} \citep[see][and references therein]{Moresco2022}.
The region of valid $k$-numbers used for the fitting, $k_{\mathrm{ini}}-k_{\mathrm{end}}$, is highlighted with vertical red lines in Fig. \ref{fig:power_spectrum} (mid-lower panel). 

Here, we summarize our method to identify the optimal $k$-wavenumber interval to estimate $P_0$, which is basically the same adopted by, e.g., \citet[][]{Cantiello2024}. For a given $E(k)$, corresponding to a specific PSF, a series of fits to eq.\,\ref{eq:power} is performed by incrementally increasing the starting wavenumber, $k_{\mathrm{start}}$, up to  user defined upper limit $k_{\mathrm{end}}$. Each iteration yields a corresponding estimate of $P_0$. The optimal wavenumber range is defined as the interval where  $P_0$  exhibits the highest stability (lower $rms$), as indicated by the dotted vertical lines in Fig. \ref{fig:power_spectrum} (bottom row, middle panel). The right-lower panel of the figure shows the fitted $P_0$ value as a function of the starting wavenumber $k_{\mathrm{start}}$ used for the fit.  
A key advantage of this approach is its efficiency and automation. 
The procedure is then  iteratively applied to each selected PSF, and the final $P_0$ value is determined as the $3 \sigma$-clipped median of the individual estimates derived for each PSF. 

\subsection{Color measurement}
\label{3.6}
The SBF signal is by definition sensitive to the properties of the  stellar populations within the galaxy. Integrated colors have been used as a proxy for these characteristics to calibrate the absolute fluctuation amplitude \citep{Tonry1990}. Thus, an accurate measurement of the color is essential for reliable  determination of $\overline{M}$ and, consequently, of the distance \citep[][]{Tonry1997,bva2001,Jensen2015,Cantiello2018}. 
Our procedure takes full advantage of the multi-band HSC-SSP and NGVS survey observations to measure integrated colors in the same annuli used for the SBF, and propagating the same masks (obtained from both bands) optimized during the fluctuation analysis stage.
Total fluxes in both bands are estimated using this mask to get the galaxy color. The MW foreground extinction correction is applied using $A_g =3.237\times E(B-V)$ and  $A_i=1.684 \times E(B-V)$ \citep{Schlafly2011}.

To quantify the color uncertainty, we propagate the contribution from the mean sky level, the instrumental readout noise and the sky background error. For each band, the uncertainty on the total flux $\sigma_{f_{\mathrm{Tot}}}^2$is computed as:
\begin{equation}
\sigma_{f_{\mathrm{Tot}}}^2=(f_{\mathrm{Tot}} + N_\mathrm{p} \times Sky) + (N_\mathrm{p}\times Sky_{\mathrm{err}})^2 + N_\mathrm{p} \times 2RON^2
\end{equation}
where $f_{\mathrm{Tot}}$ is the band-dependent total galaxy flux in the annulus, $N_\mathrm{p}$ is the number of unmasked pixel in the annulus, $RON$ is the readout noise of the instrument, $Sky$ is the mean sky level of the image, and $Sky_{\mathrm{err}}$ is the uncertainty on the mean sky level. The error on the $(g{-}i)$ color is then derived as:
\begin{equation}
    \sigma_{\mathrm{(g{-}i)}}=2.5/\log(10) \times \sqrt{\frac{\sigma_{\mathrm{f_{Tot,g}}}^2} {f_{\mathrm{Tot,g}}^2}+ \frac{\sigma_{\mathrm{{f_Tot,i}}}^2 }{f_{\mathrm{Tot,i}}^2}}
\end{equation}

\subsection{Uncertainties and quality flag}
\label{sec:quality}
Each step described above has an associated error, which then propagates in the following stages of the procedure. We treat these uncertainties as independent and, to be as consistent and conservative as possible, we add them in quadrature. These include:

\begin{itemize}

  \item Power spectrum fitting. The statistical uncertainty of the fitting procedure for eq.\,\ref{eq:power} reflects the quality of the residual image power spectrum. Typically this is a negligible error term, 
  of the order of 0.005 mag on $\Delta \overline{m}$;\newline
    
 \item PSF templates. As the measure of $\overline{m}$ is performed on different individual PSF stars (identified as “good PSF”, Sect. \ref{sec:3.3}), the statistical difference between the PSFs has to be taken into account. We use the $\sigma_{\mathrm{MAD}}$ (defined in section \ref{sec:bkg}) of the $P_\mathrm{0}$ distribution for the selected PSF references to estimate the uncertainties. This amounts for $\Delta \overline{m}\sim 0.02$ mag, on average;
 \newline
    
\item Residual power from undetected sources. 
As is usually done in the relevant literature, we adopt an empirical uncertainty of 25\% of the value of $P_\mathrm{r}$ \citep{Cantiello2007, Blakeslee2009, Jensen2021, Jensen2025}. In this work, this contribution is of the order of $\Delta \overline{m}\sim0.04$ mag;\newline

\item Background Subtraction.  This uncertainty impacts both the normalization factor in the SBF amplitude \citep{Tonry1988} and the color used in the $\overline{M}$ calibration. To estimate the effect of background uncertainty, we offset the background by $\pm 1 \sigma_{\mathrm{sky}}$ (see Sect. 3.1) and repeated the entire analysis with the new sky value. Here, the average contribution of the background uncertainty is on the order of $\Delta\overline{m}\sim 0.02$ mag. The typical impact is larger for fainter galaxies as it impacts on average for $\Delta\overline{m}\sim 0.003$ mag on bright galaxies, and $\Delta\overline{m}\sim 0.03$ mag for faint ones; \newline

\item Extinction correction. We adopted a relative error of 10\% on the extinction \citep[e.g.][]{Schlafly2011}.
\end{itemize}

In addition to propagated photometric errors, we associated a quality flag to each galaxy.
To define the confidence of the SBF analysis, we assigned a quality classification code to each $\overline{m}$ presented here, similar to the approach adopted in \citet{Cantiello2024}. The classification is assigned considering the quality of the isophotal model residuals, possible residual contamination effects of bright nearby objects, the degree of masking, the regularity of the power spectrum fitting stage, and the stability of the SBF signal measurement to small changes in the input parameters. We classify the sources as $q1$ for excellent quality, $q2$ for good quality, and $q3$ for poor quality. All bright galaxies are classified as $q1$, while the dwarf galaxies range between $q2$ and $q3$.
In the present version of FAST-SBF, the combination of these factors (e.g. quality of the residuals) is not necessarily reflected in the global error budget. In the future we plan to automatize the procedure to define $q_\mathrm{i}$ using quantitative tests on the effects described above.

\begin{figure*}
    \centering
    \includegraphics[width=1\hsize]{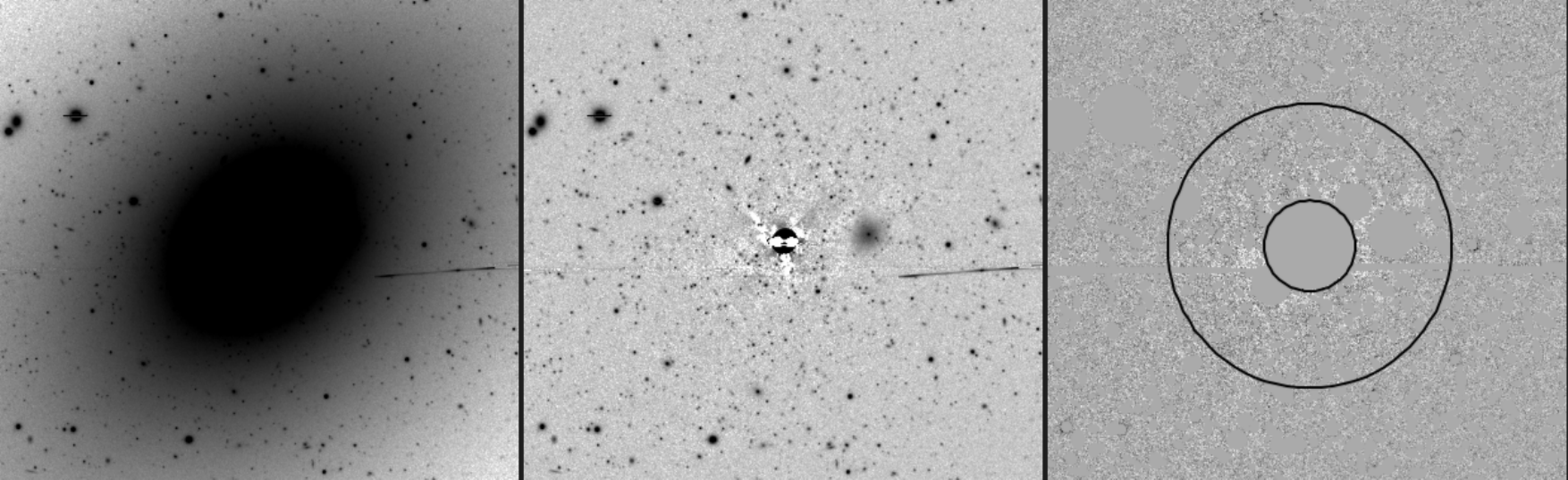}\vspace{1.5em}
    \includegraphics[width=0.45\hsize]{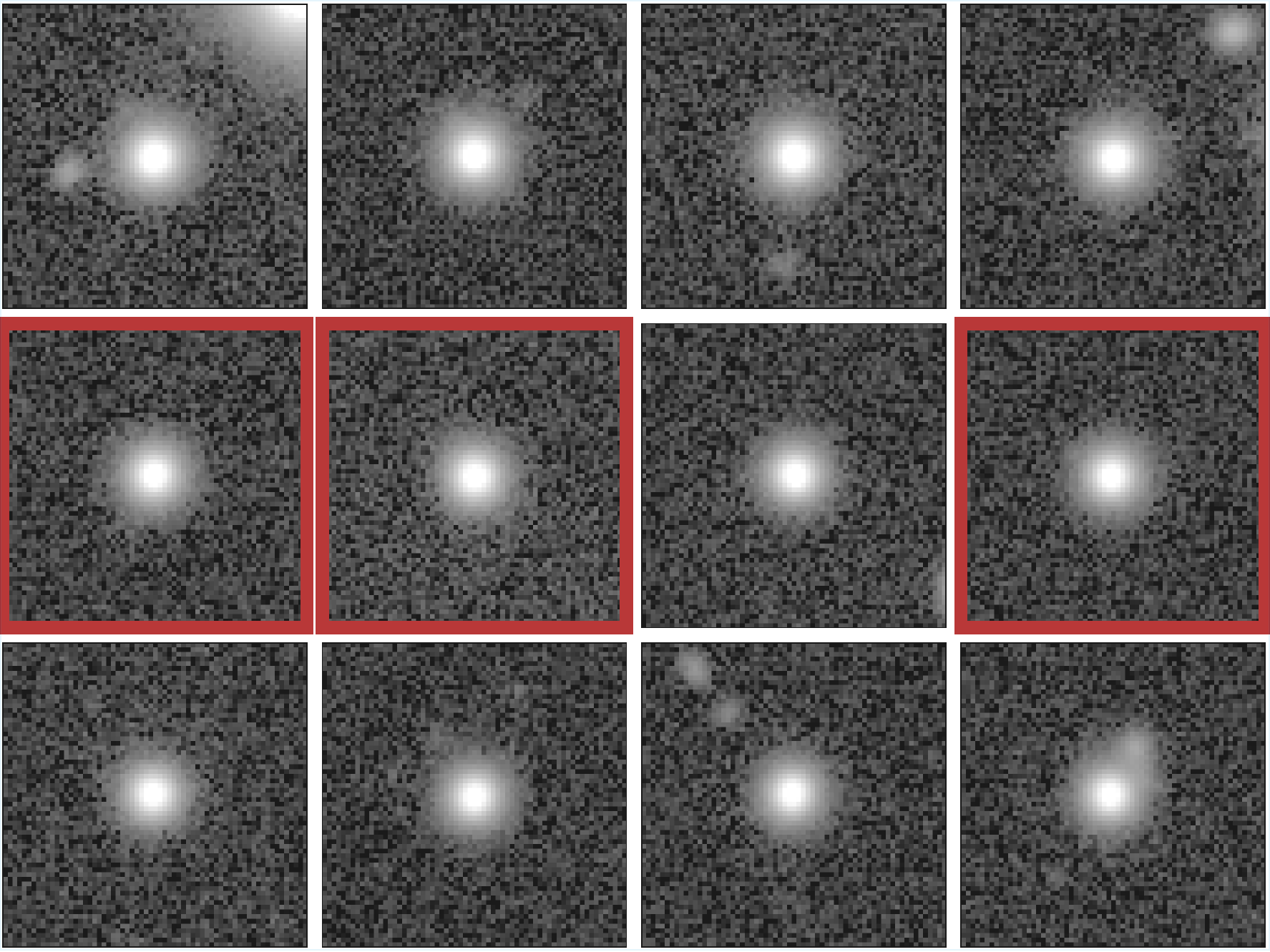}%
    \hspace{0.05\hsize}
    \includegraphics[width=0.45\hsize]{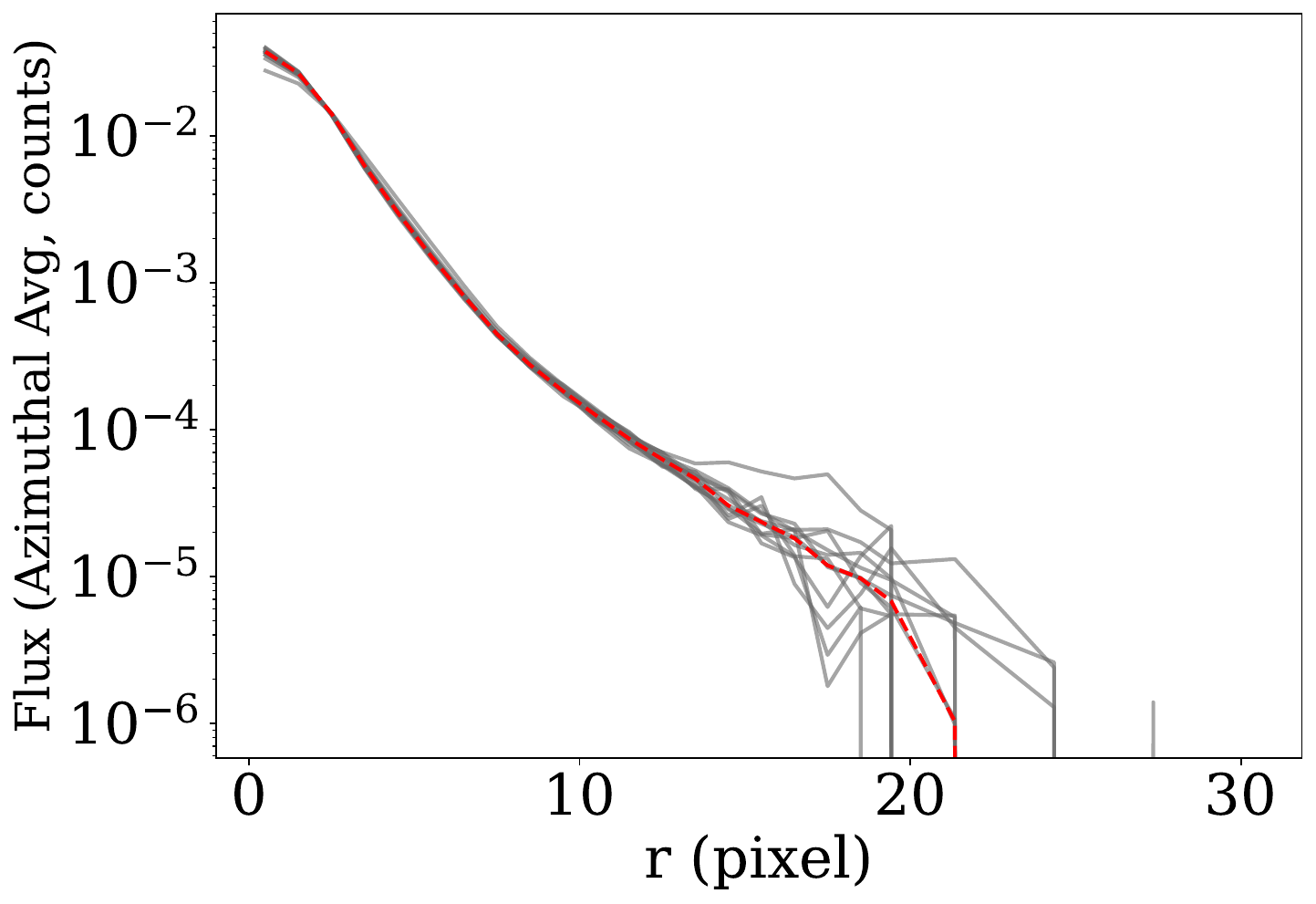}\vspace{1.5em}
    \includegraphics[width=0.333\hsize]{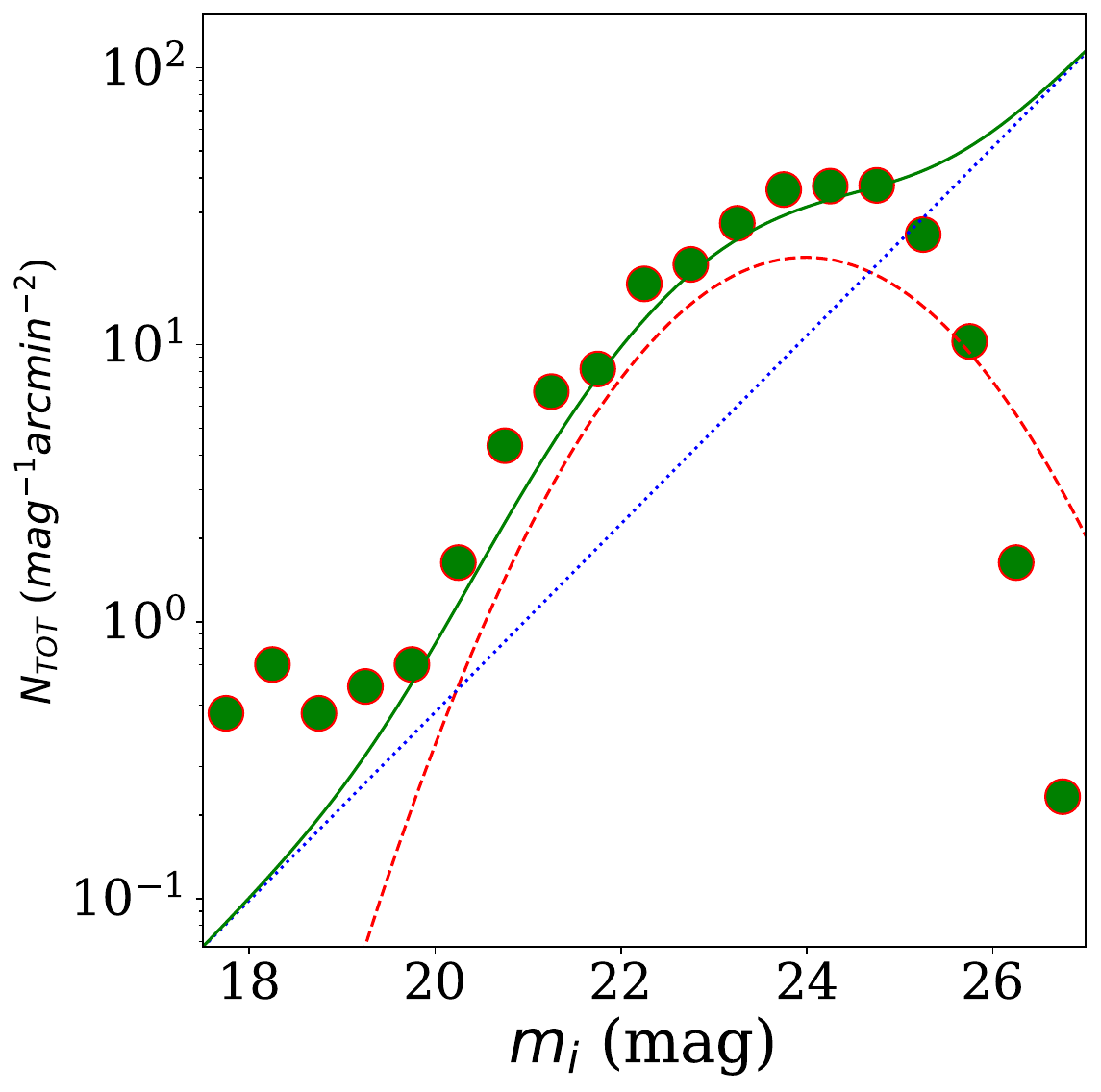}%
    \includegraphics[width=0.333\hsize]{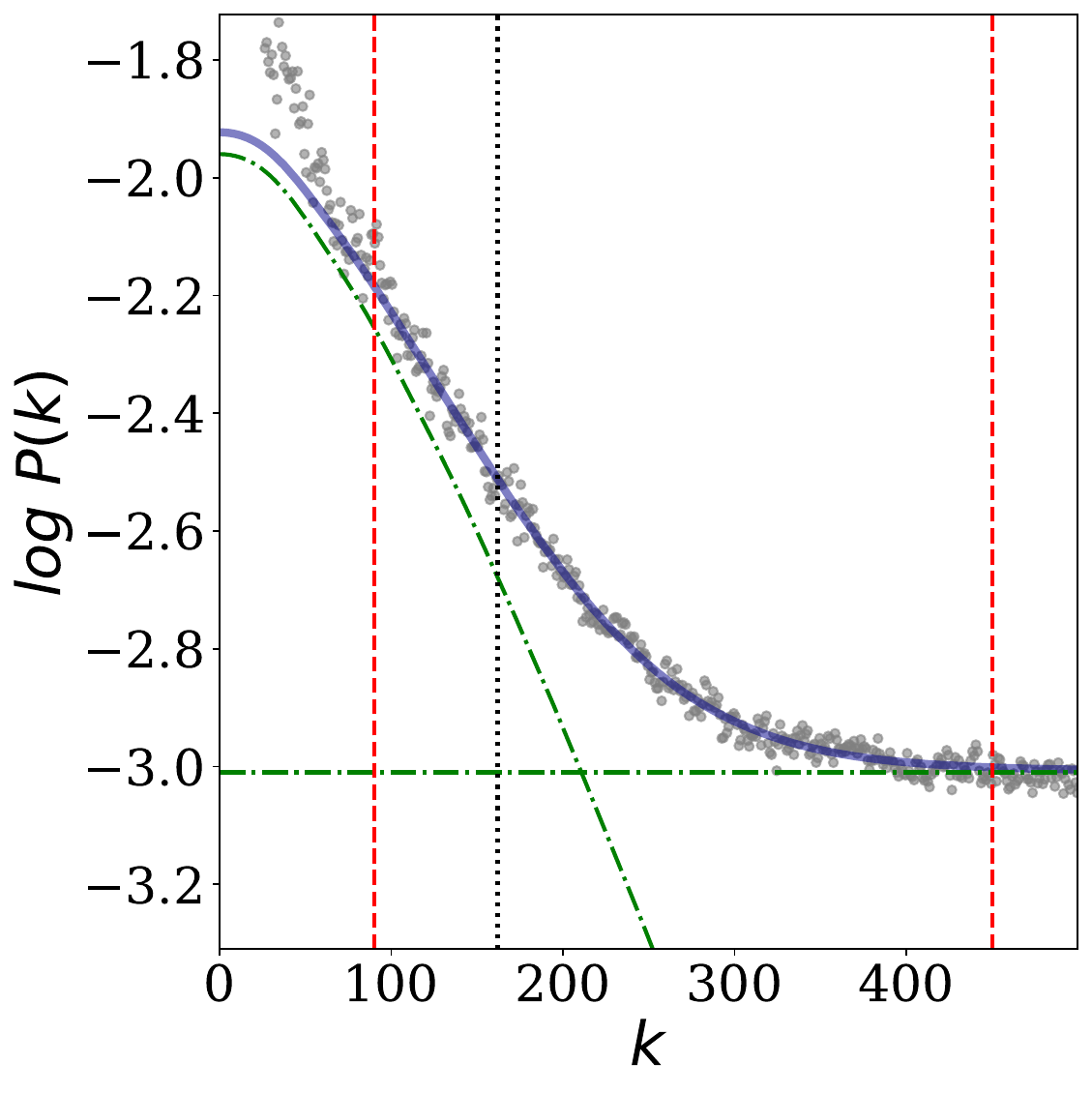}%
    \includegraphics[width=0.333\hsize]{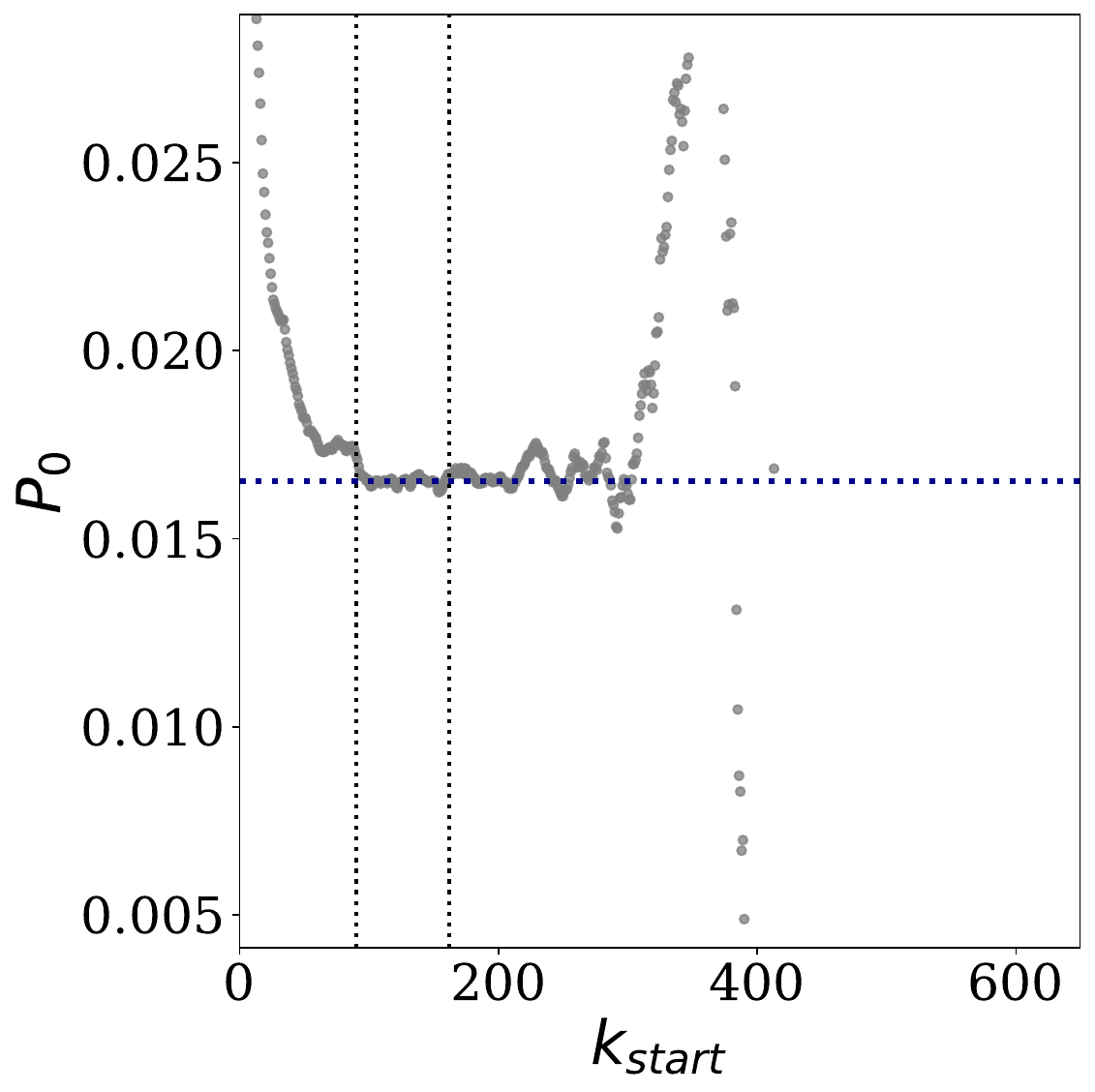}%
\caption{SBF analysis images and plots for NGC\,5813. 
Top row: $i$-band image, residual, and residual masked image (left to right). 
The black circles in the third panel show the inner and outer radii of the annulus adopted for the SBF measurements.
Middle row: example of the PSF selection process. In the left panel are shown the stars pre-selected based on photometric criteria. Stars highlighted with red borders represent a sub-sample of those subsequently chosen for the fitting.  In the right panel, the radial profile of the full sample of PSF selected is shown. The red line shows the median profile, while thin gray lines are azimuthal averaged profiles of single PSFs. The PSFs are normalized to have a total energy of one.
Bottom left panel: fitted luminosity function of external sources. The green circles represent the binned observational data. The solid green curve is the best fit to the data. The two components of the luminosity function, the background galaxies and GCLF, are shown with a blue dotted and red dashed curves, respectively. Bottom middle panel: the azimuthal average of the residual image power spectrum using one of the good PSFs. The gray dots represent the observational data. The solid blue curve is the fit obtained according to the procedure described in the text. The red dashed lines represent the range of wavenumbers used for the fit. The green dashed curve shows the PSF power spectrum. The green dotted horizontal line represents the fitted white noise value $P_1$. Bottom right panel: estimated $P_\mathrm{0}$ (grey dots) at different $k_{\mathrm{start}}$; the vertical black dotted lines indicate the most stable region in which $P_\mathrm{0}$ is calculated, while the horizontal one is the final value of $P_0$ for the specific PSF.}
    \label{fig:power_spectrum}
\end{figure*}

\subsection{Distance determination}

For SBF, to estimate the distance moduli of the measured galaxies, an estimate of the absolute SBF magnitude $\overline{M_\mathrm{i}}$ is needed. For this purpose we adopted the calibration equation from \citet{Cantiello2024} for the $(g{-}i)$ color:
\begin{equation}
\label{formula}
    \overline{M_\mathrm{i}}=a_0x^3+a_1x^2+a_2x+a_3
\end{equation}
where $a_0=10.085 \pm 6.832$, $a_1=5.20 \pm 2.042$, $a_2=2.81 \pm 0.192$ and $a_3=-1.129 \pm 0.022$, and $x=(g{-}i)-0.90$ mag represents the difference between the measured integrated color and the reference color.
This calibration is obtained in the CFHT magnitude system, which we transformed to HSC passbands using the equations derived by \citet{Kim&lee2021} from
PARSEC isochrones \citep{Bressan2012} and, like these authors, assume that the $i$-band magnitudes of the two systems
are almost identical, while for the $g$-band a correction is required.
Inverting the equation given in Sect. 5 of \citet{Kim&lee2021} we derive:
\begin{equation}
\label{eq:Kimlee}
    (g-i)_{\mathrm{CFHT}}=0.956\times[(g-i)_{\mathrm{HSC}}-0.006] 
\end{equation}
with an rms=$0.002$ mag from the {\sc{PARSEC}} isochrones \citep[see Sect. 5 of][]{Kim&lee2021}.
Table\,\ref{table:2} reports the results of the SBF analysis for the selected galaxies. 
    
\section{Analysis}
\label{sec:Results}
The primary object of this study is to introduce a new, self-contained SBF pipeline, developed to be flexible and automated, and to study the prospects for measuring SBF in the LSST by using the HSC precursor data. 
In this section, we first assess the robustness of the new pipeline on the NGVS sample, then estimate the distance to the galaxies observed by the HSC-SSP and compare the results to literature and stellar population models.  

\subsection{Comparison with independent NGVS measurements}

\begin{figure*} 
    \centering
    \includegraphics[width=1\hsize]{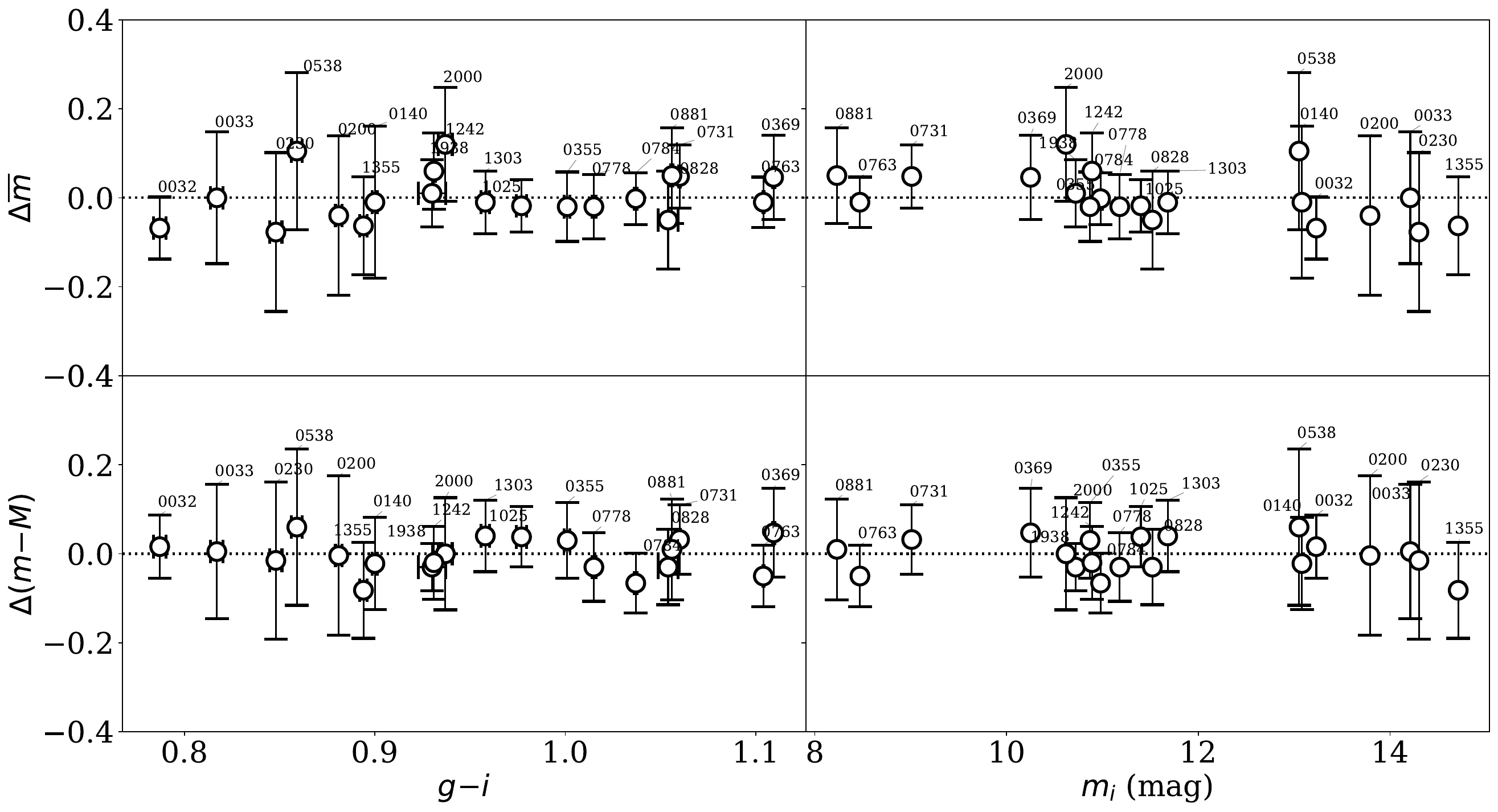}
    \caption{Comparison with the measurements based on NGVS data. The left panels show the SBF magnitude (upper panel) and distance modulus (lower panel) differences between our measurements and \citet{Cantiello2024}, as a function of the $(g{-}i)$ color. The right panels show the same comparison but as a function of the galaxy total $i$-band magnitude.}
    \label{fig:ngvs_comparison}
\end{figure*}

We compare our measurements for the SBF signal, color, and distance modulus with the results recently obtained by \cite{Cantiello2024}. The comparison is carried out for 20 galaxies in the Virgo cluster observed in the NGVS observational campaign. The galaxies were selected to cover a wide range of magnitude and color. Furthermore, since our goal is to demonstrate the consistency between the results of the two different analysis procedures, we limited the comparison to galaxies labeled with code q1 in \citet{Cantiello2024}. This was done to avoid discrepancies arising from the specific galaxy or field properties, rather than from differences in the algorithms themselves. Figure \ref{fig:ngvs_comparison} shows the difference in $\overline{m}$ and distance modulus as a function of color (left panels) and luminosity (right panels) between the two sets of measurements. The galaxies selected for this comparison span a range of color $0.7 \leq (g{-}i) \leq 1.2$ and magnitude $9 \leq m_\mathrm{i} \leq 15$ mag. The SBF measurements are consistent within the errors. Even in the few cases where the discrepancy appears relatively larger (e.g., VCC\,0538), we find excellent agreement with previous measurements when the comparison is expressed in terms of distance modulus. The difference can be attributed to minor discrepancies in the masking and the annulus size used for the measurements. These factors introduce small offsets in the $\overline{m}$ estimates but, since it also affects the color measurement, $\overline{M}$ is affected too. Hence, the final distance modulus estimates remain well within $1\sigma$ consistency.

In conclusion, the tests performed on the NGVS sample show an excellent agreement between the previous results (based on a pipeline which required heavy human intervention at all stages) and the estimates derived with the new FAST-SBF pipeline, where some modules have been successfully automated, with a net gain in speed and flexibility.

\subsection{Measurements on the HSC-SSP sample}

As discussed in Sect.\,\ref{sec:Introduction}, the SBF method is one of the few standard candles that allows measuring distances to dwarf galaxies in the local universe when single stars are not resolved. From the perspective of the forthcoming start of the LSST observations, it is realistic to expect that distances to thousands of dwarfs (or more) will be within reach. HSC-SSP data serve as a valuable testbed for such expectations, since the data reduction pipelines developed for LSST are built upon the same software framework as the HSC-SSP science pipelines. Here, we discuss the results obtained on the sample of five bright galaxies with contemporary HSC-SSP $g-$ and $i-$band data and available in the \citetalias{Tonry2001} catalog (Sect. \ref{Sec:Data}). In addition to these reference targets, we extended our measurements to a collection of visually identified dwarf galaxies with brightness profiles appropriate for SBF analysis, brighter than $m_\mathrm{B}\sim18$ mag and within two degrees of the bright galaxies. In all the figures that follow, the colors and magnitudes reported in the table have been extinction corrected.

\begin{figure*} 
    \centering
    \includegraphics[width=1\hsize]{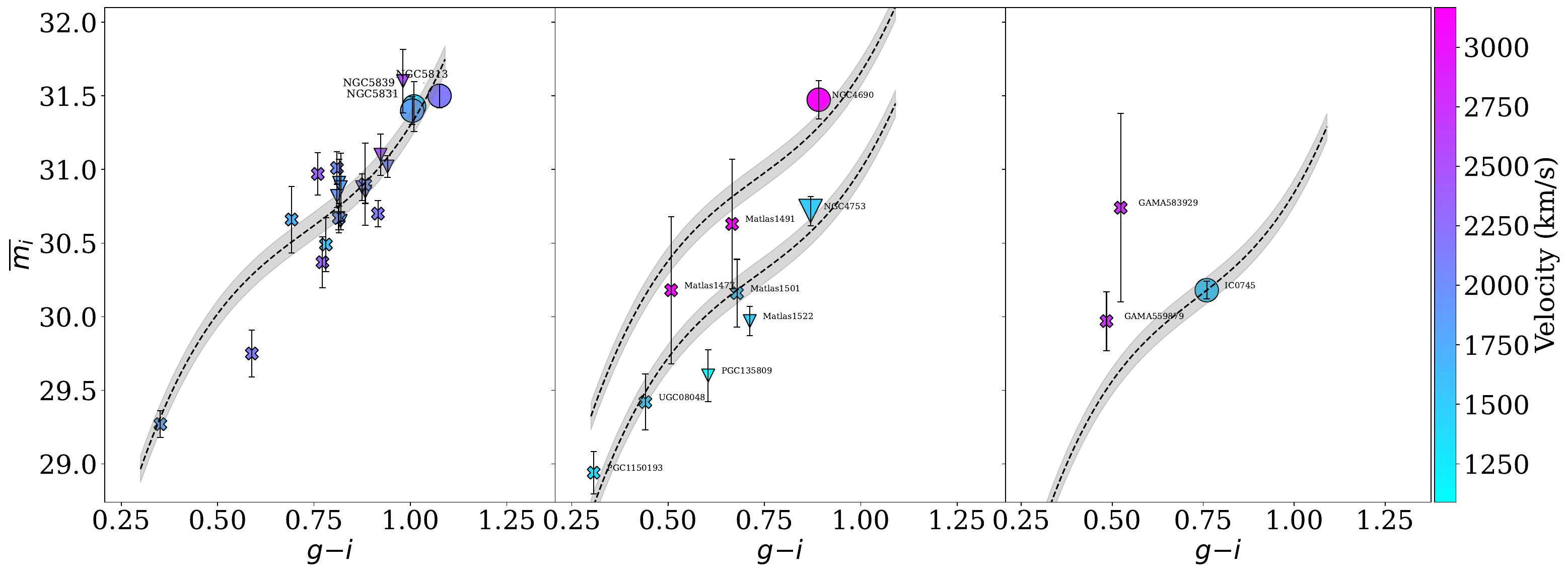}

    \caption{SBF magnitude as a function of color. Left, middle and right panels show the three different fields analyzed in this work, with the same order as in Fig. \ref{fig:spatial_distr}). The colorbar represents the velocity of the sources, $v_{\mathrm{CMB}}$. The various symbol shapes represent the assigned quality flag for the SBF measurement: circles for galaxies classified as q1, triangles for q2,  "X" for q3. Different sizes distinguish bright from dwarf galaxies. The black dashed lines represent the calibration formula from \citet{Cantiello2024} shifted as discussed in the text. The shaded grey area is the mean cosmic scatter from the relation $\sigma_{\mathrm{cos}}$, as estimated in \citet{Cantiello2024}.}
    \label{fig:mbar_col}
\end{figure*}

In Fig. \ref{fig:mbar_col} we show separately the measured SBF magnitudes as a function of galaxy $(g{-}i)$ color in the three fields. To use the calibration equation given in Eq. \ref{formula}, we converted the HSC $(g{-}i)$ colors to the NGVS system using Eq.\,\ref{eq:Kimlee}. In the figure, the color bar represents the velocity of individual galaxies relative to the CMB frame.
The black dashed line indicates the calibration relation from \cite{Cantiello2024}, shifted according to the mean distance modulus of the group, which we derived from our distances using only the confirmed group members \citep{Kourkchi2017}. Below we discuss the three fields separately.
\\
    {$\bullet$ Virgo III Cloud} - Three of the bright galaxies in our sample, NGC\,5813, NGC\,5831 and NGC\,5839, are members of NGC\,5846 group in the Virgo III cloud \citep{Tully1982}. \citet{Kourkchi2017} report that the three galaxies belong to a larger group comprising 84 members. We identified a sample of 20 dwarf galaxies which all appear, together with the bright galaxies, to lie on the expected $\overline{m}$-color relation for galaxies at the same adopted group distance. This result, combined with the recession velocities, suggest that the measured galaxies are gravitationally bound to the same group. With the exception of PGC\,1199471, all the identified galaxies are reported as group members in \citet{Kourkchi2017}. Considering only the three bright galaxies, we estimate a group distance modulus of $m-M=32.09 \pm 0.12$ corresponding to a distance of $D=26.2 \pm 1.5$ Mpc. We ignore the dwarfs to avoid any bias arising from an incomplete selection of dwarf galaxies. When the sample of dwarfs is included, we find $m-M=32.08 \pm 0.21$ (or $26.1 \pm 2.6$ Mpc).
    
    We observe the majority of dwarf galaxies in this field (left panel in the Fig.\,\ref{fig:mbar_col}) to align with the calibration relation, with an average scatter of 0.21 mag. The observed $rms$ with respect to the mean relation corresponds to an approximate depth of 0.15 mag, namely $\sim 1.8$ Mpc depth of the group, after subtracting the expected cosmic scatter \citep[$\sigma_{\mathrm{cos}}=0.09$,][]{Cantiello2024} and the median $\overline{m}$ uncertainty. We find all sources, with the exception of PGC\,1230503, MATLAS\,1944, MATLAS\,1945, and MATLAS\,2005, to be located in projection around 1.5 Mpc from the nearest reference elliptical galaxy. The remaining four dwarf galaxies, instead, are found at the projected distances of $\approx3.5$ Mpc from the main group. 
    The information derived is of paramount interest for membership confirmation for such loose, low surface brightness galaxies where spectroscopic membership typically requires prohibitive integration times, even with 8m-class telescopes.
    
\begin{figure*}
    \centering
    \includegraphics[width=1\hsize]{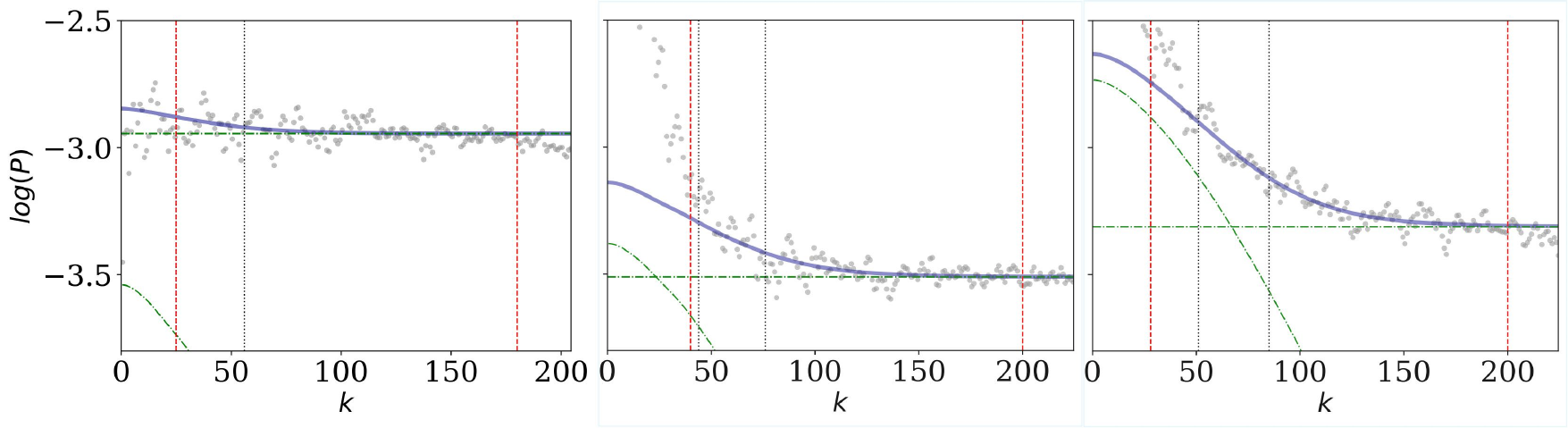}
    \caption{Comparison of power spectra measured in an empty region (left panel), for a background galaxy (PGC\,1174699, middle panel), and one of the dwarf galaxies in our sample (PGC\,1165764, right panel). The colors, points and line styles are the same as in Fig. \ref{fig:power_spectrum} (middle low panel).}
    \label{fig:dwarf_comparison}
\end{figure*}

    To test the usability of the SBF signal where the signal is low but still measurable for dwarf low surface brightness galaxies, we compared the power spectrum of one of the average dwarfs in our sample (namely PGC\,1165764) to the power spectrum of an empty field and a background galaxy. Fig. \ref{fig:dwarf_comparison} shows, from left to right, the power spectra derived over an empty region, for a background galaxy, and of PGC\,1165764, analyzed using the same configuration parameters. The background galaxy is PGC\,1174699, at $v_{\mathrm{CMB}}=12522$ km/s, and $M_\mathrm{g}\approx -20$ mag, with a distance from the Hubble law of $\approx 170$ Mpc, assuming $H_0\sim73.5$ km/s/Mpc \citep{Jensen2025}. The fluctuation signal measured for the dwarf, as observed by the relative amplitude of the PSF power spectrum (green dot-dashed lines in the panels), compared to the background galaxy and the empty field, is evidence of a stellar count fluctuation signal being detected and, therefore, of its relative closeness. For a quantitative comparison, the ratio between the fluctuation signal and the white noise component is $P_0/P_1\approx 20$ for the background galaxy PGC\,1174699, in comparison with  $P_0/P_1\approx 300$ for the dwarf galaxy PGC\,1165764. For PGC\,1174699, given its large distance, the small signal measured is attributed entirely to its undetected GC population \citep{Blakeslee1995}. Moreover, the test on the empty field indicates that the amplitude measured for the dwarf is not a bias or an artificial effect from, e.g., data processing.
    
    Before comparing our distances for the single galaxies with previous estimates, we have to consider that the \citetalias{Tonry2001} measurements were performed on a collection of data acquired in the 1990's with 2.4 to 4 m aperture telescopes and using smaller format CCDs (\citealt{Tonry1997}). 
    As already discussed in \citet{Cantiello2018}, in a few cases the distances from the first successful ground-based SBF survey could be significantly off.
    The lower resolution could, for example, result in a poor characterization of the GCLF, possibly leading to issues in the estimates of $P_\mathrm{r}$.
    
    Our distance modulus for NGC\,5813 is $\approx 0.6$ mag smaller ($\sim 30\%$ closer in distance) than \citetalias{Tonry2001}. The two estimates differ by  $\approx2.8\sigma$. The \citetalias{Tonry2001} quality flags for the galaxy (PD and Q parameters) are slightly below average. We find $m-M=31.92 \pm 0.12$ mag (namely $D= 24.2 \pm 1.3$ Mpc), which places NGC\,5813 closer to the mean group distance ($26.7 \pm 1.1$ Mpc) with respect to \citetalias{Tonry2001}. Fig. \ref{fig:power_spectrum} shows the $i$-band image of the galaxy-subtracted frame, with the annulus used for the SBF measurement. The galaxy has a large population of GCs, concentrated in the central regions, making the constraint of $P_\mathrm{r}$ crucial.
    Moreover, we also observe the presence of dust lanes around the galaxy core, which we mask before performing the SBF measurement. 
    The poorer resolution of the older \citetalias{Tonry2001} data \citep[see also Fig. 11 of][]{Cantiello2018}
    could at least in part explain the observed difference. In the case of the HSC-SSP imaging, the photometric completeness drops around $\approx 26$ mag, which is $\approx 1.5$ mag fainter than the $i$-band turnover of the expected peak of the GCLF (see Fig. \ref{fig:power_spectrum}), and is ideal for precise SBF measurement \citep{Cantiello2007, Blakeslee2009}.
    
    NGC\,5839 has several redshift independent distance estimates in the literature that we can compare with. Our distance modulus is $\approx 0.4$ larger than  \citetalias{Tonry2001}. The ground-based distance reported by
    \citetalias{Tonry2001} is superseded by the more recent estimate by \citet{Jensen2021}, based on NIR measurements from HST/WFC3 data. They report $(m-M) = 32.32 \pm 0.08$ mag, while we find $(m-M)=32.17 \pm 0.19$, with a difference of $\approx 0.7\sigma$. Another available measurement comes from the Type Ia supernova 2015bp as part of the Pantheon+SH0ES program \citep{Reiss2022}. They report $(m-M)=32.38 \pm 0.32$, within $0.55\sigma$ of our measurement. We note that the central region of NGC 5839 is contaminated by the presence of an additional source, most likely a background galaxy, making it difficult to accurately model. Moreover, \citet{Jensen2021} reported a central spiral component observed with HST F110W band, which is barely resolved in HSC $i$-band imaging. To mitigate these issues, we masked the central region when performing the SBF measurement. Unlike NGC\,5813, the sensitivity limit of the field of NGC\,5839 drops around $\approx 24.5-25$ mag. Because of this, we are able to observe $\approx 0.5-1$ mag fainter than the turnover of the GCLF, placing the galaxy at the limiting distance to have a reliable measurement of the SBF signal.
    
    For NGC\,5831, \citetalias{Tonry2001} reported a distance of $(m-M)=32.14 \pm 0.17$ mag, in perfect match with our measurement of $(m-M)=32.17 \pm 0.13$ mag. The galaxy is average in terms of \citetalias{Tonry2001} quality flags, and in the HSC-SSP image there are no visible artifacts or anomalies.
\\
    {$\bullet$ Virgo II Cloud} - In the middle panel of Fig. \ref{fig:mbar_col} we show the results for the field of NGC\,4753. As reported in \citet{Kourkchi2017}, NGC\,4753 is the brightest in a group of eight galaxies, five of which are observed in the HSC-SSP PDR2, and analyzed in this work. Moreover, within 2 deg from NGC\,4753, we identified another bright galaxy, NGC\,4690, and three fainter galaxies (namely MATLAS\,1477, MATLAS\,1491 and PGC\,135809) which presented characteristics useful for the SBF analysis.
    We find that all the galaxies reported as group members of NGC\,4753 in \citet{Kourkchi2017} follow very well the NGVS $\overline{M}$-$(g{-}i)$ calibration (Eq.\,\ref{eq:power}) shifted to the group distance, which we estimated as $m-M=31.77 \pm 0.17$ mag ($D=22.6 \pm 1.9$ Mpc) using all the confirmed members. The group had no previous mean distance measured in literature. As the majority of the galaxies analyzed here are dwarfs with relatively high SBF uncertainties, and given the limited number of members, we lack the precision necessary to constrain the depth of the group.
    PGC\,135809 shows slightly brighter $\overline{m}$ compared with the rest of the group, consistent with it being in the foreground of the group, a finding which is possibly coherent with its lower recession velocity. 

    For NGC\,4753, \citetalias{Tonry2001} reported a distance of  $(m-M) = 31.90 \pm 0.19$ mag, in excellent agreement with our estimate $(m-M) = 31.92 \pm 0.13$ mag ($D=24.3 \pm 1.5$ Mpc). The galaxy presents diffuse dust lanes extending from the center of the galaxy to the outskirts, which we masked based on the HSC-SSP $g$-band image. 
    
    In the same field, NGC\,4690, MATLAS\,1477 and MATLAS\,1491 exhibit much fainter fluctuations, consistently with their velocity $\sim$1500 km/s higher than NGC\,4753. The combined information about recession velocities and newly measured distances suggest membership to a single group in the background of NGC\,4753. Indeed, NGC\,4690 and MATLAS\,1491 are reported as part of the same group by \citet{Kourkchi2017}.
    For NGC\,4690, there are no previous distance measurements. We derive $(m-M) = 32.62 \pm 0.11$ mag, which places the galaxy at $D=33.4 \pm 1.7$ Mpc.
\\
    {$\bullet$ Leo II Cloud} - In the last panel we show the results for the field of IC\,0745, reported as member of the Leo II cloud \citep{Castignani2022}.
    The galaxy is reported as isolated in \citet{Kourkchi2017}, and indeed we visually found only 2 dwarf galaxies within 2 deg from the bright reference.
    According to our distance estimates, none of the observed dwarfs appears to be associated to IC\,0745.   
    
    For IC\,0745, \citetalias{Tonry2001} reported a quality flag largely below average. The galaxy shows strong contamination from an intermediate mass irregular galaxy at small galactocentric distance, and morphological irregularities that suggest the presence of dust patches. Comparing the $i$- and $g$-band images, we masked all regions potentially affected by dust contamination and restricted the analysis to the outskirts of the galaxy to avoid poorly modeled regions. \citetalias{Tonry2001} estimated the SBF signal in an inner annulus more centrally located than ours, covering a region possibly affected by dust extinction.
    They reported a distance modulus of $(m-M)=31.38 \pm 0.30$ mag, in comparison with our analysis which yelds $(m-M)=31.63 \pm 0.11$ mag ($D=21.2 \pm 1.1$ Mpc). The two estimates agree within $ \approx 0.8\sigma$, also owning to the relative large errors of the older measurements. 
    For the additional bright galaxy found in the field, IC\,0753, we are only complete down to $\sim0.5$ mag brighter than the turnover of the GCLF rendering the constraint on the luminosity function unreliable. Furthermore, at its distance of $\sim100$ Mpc, it is similar to the background galaxy discussed previously (PGC\,1174699), where the measured $\overline{m}$ is expected to be dominated by the contribution from the undetected GC population of the galaxy, rather than the unresolved stellar component. For these reasons, we chose to exclude IC\,0753 from the final analysis.

\subsection{Comparison with stellar population models}

In this section, we compare our measured SBF amplitudes with predictions from the Teramo-SPoT stellar population (SSP) models \citep{Cantiello2005, Raimondo2005, Raimondo2009}. The SPot-SBF models are based on a stellar population synthesis code originally described by \cite{Brocato1999}. The code is designed to accurately reproduce both the observed distribution of stars in color–magnitude diagrams and the integrated properties of star clusters and galaxies, including colors, spectral indices, and energy distributions. 
The models adopted are an updated version of the one showed in \citet{Cantiello2024}, covering a wider [Fe/H] range to include the intermediate-mass and dwarf galaxies regime.

\begin{figure}[] 
    \centering
    \includegraphics[width=1\hsize]{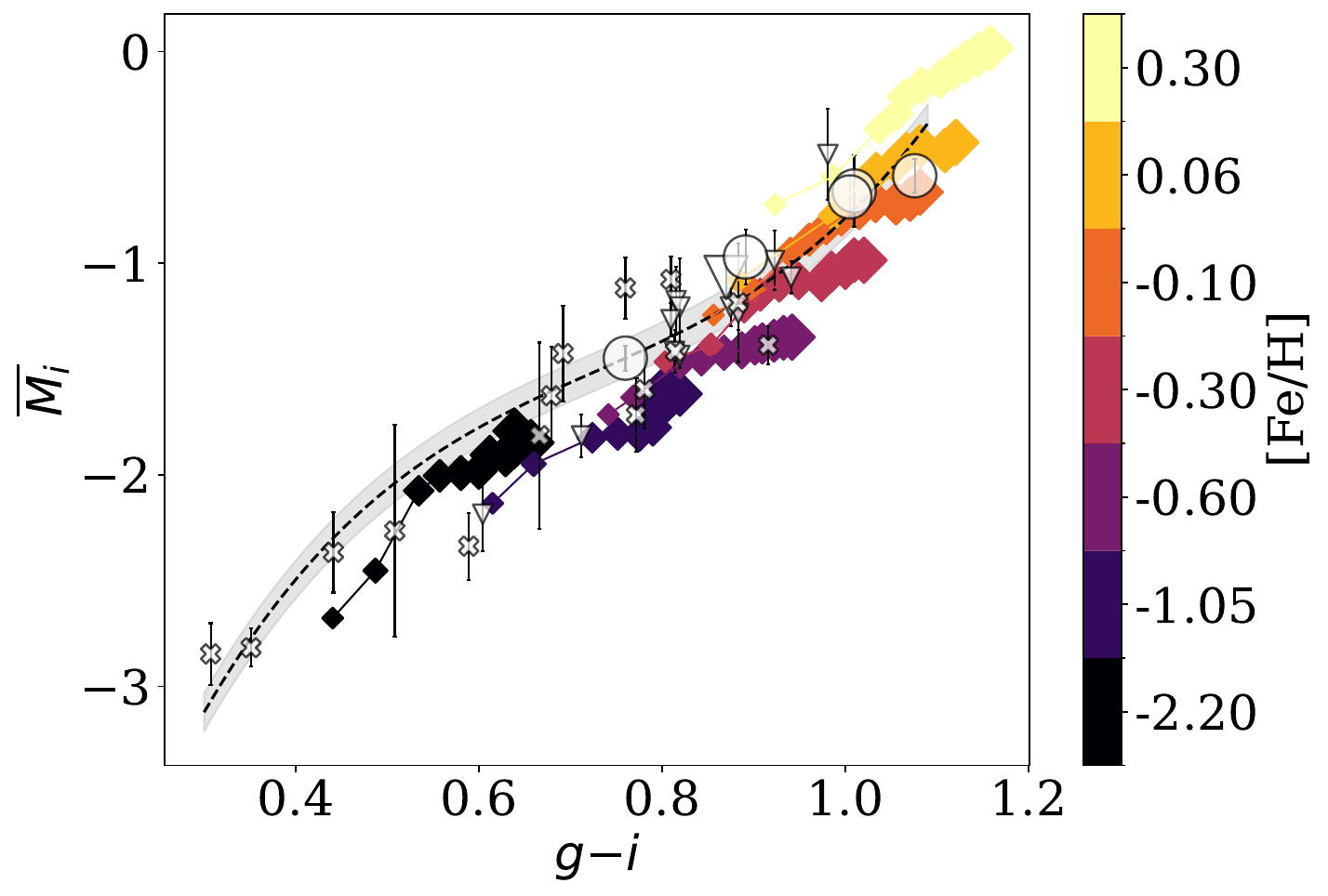}
   
    \caption{Comparison of our measurements with models. The diamonds show the SPoT SSP models for different metallicities, given by the colorbar. 
    The symbols and sizes of the galaxies are the same as in Fig. \ref{fig:mbar_col}. The black dashed line represents the calibration equation from \citet{Cantiello2024} shifted as discussed in the text. The shaded grey area is the mean cosmic scatter from the empirical calibration.}
    \label{fig:mbar_col_models}
\end{figure}

A comparison between empirical data and SSP model predictions is shown in Figure \ref{fig:mbar_col_models}. To compare the measured fluctuation magnitudes with the theoretical predictions, we shift the apparent $\overline{m}$, adopting the mean distance modulus of each group.
Most of the observed ($g{-}i$) galactic colors agree with the sequences predicted by SSP models across all metallicity ranges, for ages between 3 and 14 billion years, with two outliers at the bluest end. Note that SSP models can provide insights into the dominant stellar populations in the galaxy or in the galactic regions where SBFs were measured, while composite population models would be more appropriate for a detailed study of the stellar content of the galaxies \citep[e.g. ][]{Vazdekis2020}.

All the galaxies of the Virgo III cloud, are very well aligned with predictions. 
We find four reference galaxies (NGC\, 5813, 5831, 5839, 4753) and NGC\,4690 to be consistent with the set of models around solar metallicity, and stellar populations with ages on the order of 3--14 Gyr. As expected, the three elliptical galaxies in the Virgo III cloud (NGC\,5813, NGC\,5839, NGC\,5831) cover the region for old stellar population age. NGC\,4753 and NGC\,4690 instead, are aligned with ages of around 3--6 Gyr, which is not unexpected considering their relatively irregular morphology. IC\,0745 shows sign of star-formation in the very center, which can explain its blue color and position in the region of models having sub-solar metallicity and young ages.
On the other hand, most of the dwarf galaxies in the sample align with SSP models having [Fe/H] $= -0.10$ dex and lower, and with ages ranging from 3 to 8 Gyr, in agreement with previous studies suggesting that such systems generally host relatively old (5–10 Gyr) and subsolar-metallicity stellar populations \citep{Kirby2013,Weisz2014}.

In conclusion, the agreement between measures and predictions for the appropriate range of age and metallicity provides further support to the reliability of the results presented..

\section{Summary}
\label{Sec:Conclusions}

In this work, we introduced FAST-SBF, a new, flexible Python-based pipeline designed for SBF measurements. The code has been developed for upcoming next-generation of large astronomical surveys (particularly LSST, Euclid, and the Roman Space Telescope), aiming to build a tool that is fully automated and able to handle large datasets.
In this work, we tested the procedure on ground-based imaging data from the HSC-SSP survey, an LSST precursor, taking as reference five galaxies already studied in the literature. Starting from these, we expanded our analysis to include both bright and dwarf galaxies in the same fields. In general, we obtained accurate measurements for the distances of the galaxies already known in the literature, reporting results typically in good agreement with the previous measures, confirming the robustness of the new pipeline presented here. Moreover, we extended the analysis to dwarf galaxies lacking prior distance estimates, and used their consistency with the empirical SBF–color relation to assess group membership.

Further validation was carried out using measures from the more recent NGVS program, where we compared our results with the estimates from \citet{Cantiello2024} for 20 galaxies. Again, we found excellent consistency across SBF magnitudes, colors, and derived distances. The measurements also show good agreement with predictions from stellar population synthesis models, particularly in the age and metallicity range expected for early-type galaxies and dwarfs in the local universe.

Overall, the results demonstrate the reliability of the new procedure and its potential for enabling precise, automated distance measurements across a wide variety of galaxy types. Also, the procedure has allowed us to assess the group membership of low-luminosity dwarf galaxies. 

The current version of FAST-SBF is more automated and portable than previous implementations of SBF measurement pipelines. However, the code is not yet ready for public release. At this stage, obtaining reliable results requires specific training to understand the caveats and to correctly handle those modules that demand particular attention (e.g., contamination from residual sources and proper power-spectrum fitting). While the code can produce fluctuation magnitudes for any target, the scientific validity of these measurements depends critically on both a thorough understanding of the data and the correct execution of the analysis steps.
Researchers interested in using FAST-SBF are encouraged to contact the second author (at \href{mailto:michele.cantiello@inaf.it}{michele.cantiello@inaf.it}) to discuss access to the code and receive the necessary training for its appropriate use.

\begin{acknowledgements}
GR acknowledges support from the INAF “Astrofisica Fondamentale” Mini-grant 2024 n.17/RSN1 (PI G. Riccio). MC acknowledges support from ASI–INAF grant no. 2024-10-HH.0 (WP8420), the ESO Scientific Visitor Programme, and INAF GO-grant no. 12/2024 (P.I. M. Cantiello). RH and MC acknowledge support from the project “INAF-EDGE” (Large Grant 12-2022, P.I. L. Hunt). NH acknowledges support from the Polish National Science Centre grant 2023/50/A/ST9/00579. MM acknowledges support from the ESO Studentship program. This research has made use of the NASA/IPAC Extra- galactic Database (NED\footnote{https://ned.ipac.caltech.edu/}), which is funded by the National Aeronautics and Space Administration and operated by the California Institute of Technology. \textit{Software}: Numpy \citep{harris2020array}, Scipy \citep{2020SciPy-NMeth}, Source Extractor \citep{Bertin1996}. We are grateful to Adriano Pietrinferni and Santi Cassisi for providing us tables with key points of their evolutionary tracks published in \citet{Hidalgo2018}, and \citet{Adriano2021}.

\end{acknowledgements}

\bibliographystyle{aa} 
\bibliography{aa56567-25}

\begin{thebibliography}{60}
\expandafter\ifx\csname natexlab\endcsname\relax\def\natexlab#1{#1}\fi

\bibitem[{{Aaronson} {et~al.}(1982){Aaronson}, {Huchra}, {Mould}, {Schechter}, \& {Tully}}]{Tully1982}
{Aaronson}, M., {Huchra}, J., {Mould}, J., {Schechter}, P.~L., \& {Tully}, R.~B. 1982, \apj, 258, 64

\bibitem[{{Aihara} {et~al.}(2019){Aihara}, {AlSayyad}, {Ando}, {Armstrong}, {Bosch}, {Egami}, {Furusawa}, {Furusawa}, {Goulding}, {Harikane}, {Hikage}, {Ho}, {Hsieh}, {Huang}, {Ikeda}, {Imanishi}, {Ito}, {Iwata}, {Jaelani}, {Kakuma}, {Kawana}, {Kikuta}, {Kobayashi}, {Koike}, {Komiyama}, {Li}, {Liang}, {Lin}, {Luo}, {Lupton}, {Lust}, {MacArthur}, {Matsuoka}, {Mineo}, {Miyatake}, {Miyazaki}, {More}, {Murata}, {Namiki}, {Nishizawa}, {Oguri}, {Okabe}, {Okamoto}, {Okura}, {Ono}, {Onodera}, {Onoue}, {Osato}, {Ouchi}, {Shibuya}, {Strauss}, {Sugiyama}, {Suto}, {Takada}, {Takagi}, {Takata}, {Takita}, {Tanaka}, {Terai}, {Toba}, {Uchiyama}, {Utsumi}, {Wang}, {Wang}, \& {Yamada}}]{Aihara2019}
{Aihara}, H., {AlSayyad}, Y., {Ando}, M., {et~al.} 2019, \pasj, 71, 114

\bibitem[{{Aihara} {et~al.}(2022){Aihara}, {AlSayyad}, {Ando}, {Armstrong}, {Bosch}, {Egami}, {Furusawa}, {Furusawa}, {Harasawa}, {Harikane}, {Hsieh}, {Ikeda}, {Ito}, {Iwata}, {Kodama}, {Koike}, {Kokubo}, {Komiyama}, {Li}, {Liang}, {Lin}, {Lupton}, {Lust}, {MacArthur}, {Mawatari}, {Mineo}, {Miyatake}, {Miyazaki}, {More}, {Morishima}, {Murayama}, {Nakajima}, {Nakata}, {Nishizawa}, {Oguri}, {Okabe}, {Okura}, {Ono}, {Osato}, {Ouchi}, {Pan}, {Plazas Malag{\'o}n}, {Price}, {Reed}, {Rykoff}, {Shibuya}, {Simunovic}, {Strauss}, {Sugimori}, {Suto}, {Suzuki}, {Takada}, {Takagi}, {Takata}, {Takita}, {Tanaka}, {Tang}, {Taranu}, {Terai}, {Toba}, {Turner}, {Uchiyama}, {Vijarnwannaluk}, {Waters}, {Yamada}, {Yamamoto}, \& {Yamashita}}]{Aiharapdr3}
{Aihara}, H., {AlSayyad}, Y., {Ando}, M., {et~al.} 2022, \pasj, 74, 247

\bibitem[{{Bernstein} \& {Jarvis}(2002)}]{Bernstein2002}
{Bernstein}, G.~M. \& {Jarvis}, M. 2002, \aj, 123, 583

\bibitem[{{Bertin} \& {Arnouts}(1996)}]{Bertin1996}
{Bertin}, E. \& {Arnouts}, S. 1996, \aaps, 117, 393

\bibitem[{{Blakeslee} {et~al.}(1999){Blakeslee}, {Ajhar}, \& {Tonry}}]{Blakeslee1999}
{Blakeslee}, J.~P., {Ajhar}, E.~A., \& {Tonry}, J.~L. 1999, in Astrophysics and Space Science Library, Vol. 237, Post-Hipparcos Cosmic Candles, ed. A.~{Heck} \& F.~{Caputo}, 181

\bibitem[{{Blakeslee} {et~al.}(2021){Blakeslee}, {Jensen}, {Ma}, {Milne}, \& {Greene}}]{Blakeslee2021}
{Blakeslee}, J.~P., {Jensen}, J.~B., {Ma}, C.-P., {Milne}, P.~A., \& {Greene}, J.~E. 2021, \apj, 911, 65

\bibitem[{{Blakeslee} {et~al.}(2009){Blakeslee}, {Jord{\'a}n}, {Mei}, {C{\^o}t{\'e}}, {Ferrarese}, {Infante}, {Peng}, {Tonry}, \& {West}}]{Blakeslee2009}
{Blakeslee}, J.~P., {Jord{\'a}n}, A., {Mei}, S., {et~al.} 2009, \apj, 694, 556

\bibitem[{{Blakeslee} \& {Tonry}(1995)}]{Blakeslee1995}
{Blakeslee}, J.~P. \& {Tonry}, J.~L. 1995, \apj, 442, 579

\bibitem[{{Blakeslee} {et~al.}(2001){Blakeslee}, {Vazdekis}, \& {Ajhar}}]{bva2001}
{Blakeslee}, J.~P., {Vazdekis}, A., \& {Ajhar}, E.~A. 2001, \mnras, 320, 193

\bibitem[{{Bressan} {et~al.}(2012){Bressan}, {Marigo}, {Girardi}, {Salasnich}, {Dal Cero}, {Rubele}, \& {Nanni}}]{Bressan2012}
{Bressan}, A., {Marigo}, P., {Girardi}, L., {et~al.} 2012, \mnras, 427, 127

\bibitem[{{Brocato} {et~al.}(1999){Brocato}, {Castellani}, {Raimondo}, \& {Romaniello}}]{Brocato1999}
{Brocato}, E., {Castellani}, V., {Raimondo}, G., \& {Romaniello}, M. 1999, \aaps, 136, 65

\bibitem[{{Buzzoni}(1993)}]{Buzzoni1993}
{Buzzoni}, A. 1993, \aap, 275, 433

\bibitem[{{Cantiello} {et~al.}(2007){Cantiello}, {Blakeslee}, {Raimondo}, {Brocato}, \& {Capaccioli}}]{Cantiello2007}
{Cantiello}, M., {Blakeslee}, J., {Raimondo}, G., {Brocato}, E., \& {Capaccioli}, M. 2007, \apj, 668, 130

\bibitem[{{Cantiello} \& {Blakeslee}(2023)}]{Cantielloreview}
{Cantiello}, M. \& {Blakeslee}, J.~P. 2023, arXiv e-prints, arXiv:2307.03116

\bibitem[{{Cantiello} {et~al.}(2024){Cantiello}, {Blakeslee}, {Ferrarese}, {C{\^o}t{\'e}}, {Raimondo}, {Cuillandre}, {Durrell}, {Gwyn}, {Hazra}, {Peng}, {Roediger}, {S{\'a}nchez-Janssen}, \& {Kurzner}}]{Cantiello2024}
{Cantiello}, M., {Blakeslee}, J.~P., {Ferrarese}, L., {et~al.} 2024, \apj, 966, 145

\bibitem[{{Cantiello} {et~al.}(2018){Cantiello}, {Blakeslee}, {Ferrarese}, {C{\^o}t{\'e}}, {Roediger}, {Raimondo}, {Peng}, {Gwyn}, {Durrell}, \& {Cuillandre}}]{Cantiello2018}
{Cantiello}, M., {Blakeslee}, J.~P., {Ferrarese}, L., {et~al.} 2018, \apj, 856, 126

\bibitem[{{Cantiello} {et~al.}(2005){Cantiello}, {Blakeslee}, {Raimondo}, {Mei}, {Brocato}, \& {Capaccioli}}]{Cantiello2005}
{Cantiello}, M., {Blakeslee}, J.~P., {Raimondo}, G., {et~al.} 2005, \apj, 634, 239

\bibitem[{{Cantiello} {et~al.}(2003){Cantiello}, {Raimondo}, {Brocato}, \& {Capaccioli}}]{Cantiello2003}
{Cantiello}, M., {Raimondo}, G., {Brocato}, E., \& {Capaccioli}, M. 2003, \aj, 125, 2783

\bibitem[{{Castignani} {et~al.}(2022){Castignani}, {Vulcani}, {Finn}, {Combes}, {Jablonka}, {Rudnick}, {Zaritsky}, {Whalen}, {Conger}, {De Lucia}, {Desai}, {Koopmann}, {Moustakas}, {Norman}, \& {Townsend}}]{Castignani2022}
{Castignani}, G., {Vulcani}, B., {Finn}, R.~A., {et~al.} 2022, \apjs, 259, 43

\bibitem[{{Duc} {et~al.}(2015){Duc}, {Cuillandre}, {Karabal}, {Cappellari}, {Alatalo}, {Blitz}, {Bournaud}, {Bureau}, {Crocker}, {Davies}, {Davis}, {de Zeeuw}, {Emsellem}, {Khochfar}, {Krajnovi{\'c}}, {Kuntschner}, {McDermid}, {Michel-Dansac}, {Morganti}, {Naab}, {Oosterloo}, {Paudel}, {Sarzi}, {Scott}, {Serra}, {Weijmans}, \& {Young}}]{Duc2015}
{Duc}, P.-A., {Cuillandre}, J.-C., {Karabal}, E., {et~al.} 2015, \mnras, 446, 120

\bibitem[{{Durrell} {et~al.}(2014){Durrell}, {C{\^o}t{\'e}}, {Peng}, {Blakeslee}, {Ferrarese}, {Mihos}, {Puzia}, {Lan{\c{c}}on}, {Liu}, {Zhang}, {Cuillandre}, {McConnachie}, {Jord{\'a}n}, {Accetta}, {Boissier}, {Boselli}, {Courteau}, {Duc}, {Emsellem}, {Gwyn}, {Mei}, \& {Taylor}}]{Durrell2014}
{Durrell}, P.~R., {C{\^o}t{\'e}}, P., {Peng}, E.~W., {et~al.} 2014, \apj, 794, 103

\bibitem[{{Ferrarese} {et~al.}(2012){Ferrarese}, {C{\^o}t{\'e}}, {Cuillandre}, {Gwyn}, {Peng}, {MacArthur}, {Duc}, {Boselli}, {Mei}, {Erben}, {McConnachie}, {Durrell}, {Mihos}, {Jord{\'a}n}, {Lan{\c{c}}on}, {Puzia}, {Emsellem}, {Balogh}, {Blakeslee}, {van Waerbeke}, {Gavazzi}, {Vollmer}, {Kavelaars}, {Woods}, {Ball}, {Boissier}, {Courteau}, {Ferriere}, {Gavazzi}, {Hildebrandt}, {Hudelot}, {Huertas-Company}, {Liu}, {McLaughlin}, {Mellier}, {Milkeraitis}, {Schade}, {Balkowski}, {Bournaud}, {Carlberg}, {Chapman}, {Hoekstra}, {Peng}, {Sawicki}, {Simard}, {Taylor}, {Tully}, {van Driel}, {Wilson}, {Burdullis}, {Mahoney}, \& {Manset}}]{Ferrarese2012}
{Ferrarese}, L., {C{\^o}t{\'e}}, P., {Cuillandre}, J.-C., {et~al.} 2012, \apjs, 200, 4

\bibitem[{Fischler \& Bolles(1981)}]{RANSAC}
Fischler, M.~A. \& Bolles, R.~C. 1981, Commun. ACM, 24, 381–395

\bibitem[{{Habas} {et~al.}(2020){Habas}, {Marleau}, {Duc}, {Durrell}, {Paudel}, {Poulain}, {S{\'a}nchez-Janssen}, {Sreejith}, {Ramasawmy}, {Stemock}, {Leach}, {Cuillandre}, {Gwyn}, {Agnello}, {B{\'\i}lek}, {Fensch}, {M{\"u}ller}, {Peng}, \& {van der Burg}}]{Habas2020}
{Habas}, R., {Marleau}, F.~R., {Duc}, P.-A., {et~al.} 2020, \mnras, 491, 1901

\bibitem[{Harris {et~al.}(2020)Harris, Millman, van~der Walt, Gommers, Virtanen, Cournapeau, Wieser, Taylor, Berg, Smith, Kern, Picus, Hoyer, van Kerkwijk, Brett, Haldane, del R{\'{i}}o, Wiebe, Peterson, G{\'{e}}rard-Marchant, Sheppard, Reddy, Weckesser, Abbasi, Gohlke, \& Oliphant}]{harris2020array}
Harris, C.~R., Millman, K.~J., van~der Walt, S.~J., {et~al.} 2020, Nature, 585, 357

\bibitem[{{Harris} {et~al.}(2002){Harris}, {Harris}, {Holland}, \& {McLaughlin}}]{Harris2002}
{Harris}, W.~E., {Harris}, G. L.~H., {Holland}, S.~T., \& {McLaughlin}, D.~E. 2002, \aj, 124, 1435

\bibitem[{{Hidalgo} {et~al.}(2018){Hidalgo}, {Pietrinferni}, {Cassisi}, {Salaris}, {Mucciarelli}, {Savino}, {Aparicio}, {Silva Aguirre}, \& {Verma}}]{Hidalgo2018}
{Hidalgo}, S.~L., {Pietrinferni}, A., {Cassisi}, S., {et~al.} 2018, \apj, 856, 125

\bibitem[{{Ivezi{\'c}} {et~al.}(2019){Ivezi{\'c}}, {Kahn}, {Tyson}, {Abel}, {Acosta}, {Allsman}, {Alonso}, {AlSayyad}, {Anderson}, {Andrew}, {Angel}, {Angeli}, {Ansari}, {Antilogus}, {Araujo}, {Armstrong}, {Arndt}, {Astier}, {Aubourg}, {Auza}, {Axelrod}, {Bard}, {Barr}, {Barrau}, {Bartlett}, {Bauer}, {Bauman}, {Baumont}, {Bechtol}, {Bechtol}, {Becker}, {Becla}, {Beldica}, {Bellavia}, {Bianco}, {Biswas}, {Blanc}, {Blazek}, {Bland ford}, {Bloom}, {Bogart}, {Bond}, {Booth}, {Borgland}, {Borne}, {Bosch}, {Boutigny}, {Brackett}, {Bradshaw}, {Brand t}, {Brown}, {Bullock}, {Burchat}, {Burke}, {Cagnoli}, {Calabrese}, {Callahan}, {Callen}, {Carlin}, {Carlson}, {Chand rasekharan}, {Charles-Emerson}, {Chesley}, {Cheu}, {Chiang}, {Chiang}, {Chirino}, {Chow}, {Ciardi}, {Claver}, {Cohen-Tanugi}, {Cockrum}, {Coles}, {Connolly}, {Cook}, {Cooray}, {Covey}, {Cribbs}, {Cui}, {Cutri}, {Daly}, {Daniel}, {Daruich}, {Daubard}, {Daues}, {Dawson}, {Delgado}, {Dellapenna}, {de Peyster}, {de Val-Borro}, {Digel}, {Doherty}, {Dubois},
  {Dubois-Felsmann}, {Durech}, {Economou}, {Eifler}, {Eracleous}, {Emmons}, {Fausti Neto}, {Ferguson}, {Figueroa}, {Fisher-Levine}, {Focke}, {Foss}, {Frank}, {Freemon}, {Gangler}, {Gawiser}, {Geary}, {Gee}, {Geha}, {Gessner}, {Gibson}, {Gilmore}, {Glanzman}, {Glick}, {Goldina}, {Goldstein}, {Goodenow}, {Graham}, {Gressler}, {Gris}, {Guy}, {Guyonnet}, {Haller}, {Harris}, {Hascall}, {Haupt}, {Hernand ez}, {Herrmann}, {Hileman}, {Hoblitt}, {Hodgson}, {Hogan}, {Howard}, {Huang}, {Huffer}, {Ingraham}, {Innes}, {Jacoby}, {Jain}, {Jammes}, {Jee}, {Jenness}, {Jernigan}, {Jevremovi{\'c}}, {Johns}, {Johnson}, {Johnson}, {Jones}, {Juramy-Gilles}, {Juri{\'c}}, {Kalirai}, {Kallivayalil}, {Kalmbach}, {Kantor}, {Karst}, {Kasliwal}, {Kelly}, {Kessler}, {Kinnison}, {Kirkby}, {Knox}, {Kotov}, {Krabbendam}, {Krughoff}, {Kub{\'a}nek}, {Kuczewski}, {Kulkarni}, {Ku}, {Kurita}, {Lage}, {Lambert}, {Lange}, {Langton}, {Le Guillou}, {Levine}, {Liang}, {Lim}, {Lintott}, {Long}, {Lopez}, {Lotz}, {Lupton}, {Lust}, {MacArthur}, {Mahabal},
  {Mand elbaum}, {Markiewicz}, {Marsh}, {Marshall}, {Marshall}, {May}, {McKercher}, {McQueen}, {Meyers}, {Migliore}, {Miller}, {Mills}, {Miraval}, {Moeyens}, {Moolekamp}, {Monet}, {Moniez}, {Monkewitz}, {Montgomery}, {Morrison}, {Mueller}, {Muller}, {Mu{\~n}oz Arancibia}, {Neill}, {Newbry}, {Nief}, {Nomerotski}, {Nordby}, {O'Connor}, {Oliver}, {Olivier}, {Olsen}, {O'Mullane}, {Ortiz}, {Osier}, {Owen}, {Pain}, {Palecek}, {Parejko}, {Parsons}, {Pease}, {Peterson}, {Peterson}, {Petravick}, {Libby Petrick}, {Petry}, {Pierfederici}, {Pietrowicz}, {Pike}, {Pinto}, {Plante}, {Plate}, {Plutchak}, {Price}, {Prouza}, {Radeka}, {Rajagopal}, {Rasmussen}, {Regnault}, {Reil}, {Reiss}, {Reuter}, {Ridgway}, {Riot}, {Ritz}, {Robinson}, {Roby}, {Roodman}, {Rosing}, {Roucelle}, {Rumore}, {Russo}, {Saha}, {Sassolas}, {Schalk}, {Schellart}, {Schindler}, {Schmidt}, {Schneider}, {Schneider}, {Schoening}, {Schumacher}, {Schwamb}, {Sebag}, {Selvy}, {Sembroski}, {Seppala}, {Serio}, {Serrano}, {Shaw}, {Shipsey}, {Sick}, {Silvestri},
  {Slater}, {Smith}, {Smith}, {Sobhani}, {Soldahl}, {Storrie-Lombardi}, {Stover}, {Strauss}, {Street}, {Stubbs}, {Sullivan}, {Sweeney}, {Swinbank}, {Szalay}, {Takacs}, {Tether}, {Thaler}, {Thayer}, {Thomas}, {Thornton}, {Thukral}, {Tice}, {Trilling}, {Turri}, {Van Berg}, {Vanden Berk}, {Vetter}, {Virieux}, {Vucina}, {Wahl}, {Walkowicz}, {Walsh}, {Walter}, {Wang}, {Wang}, {Warner}, {Wiecha}, {Willman}, {Winters}, {Wittman}, {Wolff}, {Wood-Vasey}, {Wu}, {Xin}, {Yoachim}, \& {Zhan}}]{Ivezic2019}
{Ivezi{\'c}}, {\v{Z}}., {Kahn}, S.~M., {Tyson}, J.~A., {et~al.} 2019, \apj, 873, 111

\bibitem[{{Jedrzejewski}(1987)}]{Jedrzejewski1987}
{Jedrzejewski}, R.~I. 1987, \mnras, 226, 747

\bibitem[{{Jensen} {et~al.}(2025){Jensen}, {Blakeslee}, {Cantiello}, {Cowles}, {Anand}, {Tully}, {Kourkchi}, \& {Raimondo}}]{Jensen2025}
{Jensen}, J.~B., {Blakeslee}, J.~P., {Cantiello}, M., {et~al.} 2025, \apj, 987, 87

\bibitem[{{Jensen} {et~al.}(2015){Jensen}, {Blakeslee}, {Gibson}, {Lee}, {Cantiello}, {Raimondo}, {Boyer}, \& {Cho}}]{Jensen2015}
{Jensen}, J.~B., {Blakeslee}, J.~P., {Gibson}, Z., {et~al.} 2015, \apj, 808, 91

\bibitem[{{Jensen} {et~al.}(2021){Jensen}, {Blakeslee}, {Ma}, {Milne}, {Brown}, {Cantiello}, {Garnavich}, {Greene}, {Lucey}, {Phan}, {Tully}, \& {Wood}}]{Jensen2021}
{Jensen}, J.~B., {Blakeslee}, J.~P., {Ma}, C.-P., {et~al.} 2021, \apjs, 255, 21

\bibitem[{{Jord{\'a}n} {et~al.}(2004){Jord{\'a}n}, {Blakeslee}, {Peng}, {Mei}, {C{\^o}t{\'e}}, {Ferrarese}, {Tonry}, {Merritt}, {Milosavljevi{\'c}}, \& {West}}]{Jordan2004}
{Jord{\'a}n}, A., {Blakeslee}, J.~P., {Peng}, E.~W., {et~al.} 2004, \apjs, 154, 509

\bibitem[{{Jord{\'a}n} {et~al.}(2009){Jord{\'a}n}, {Peng}, {Blakeslee}, {C{\^o}t{\'e}}, {Eyheramendy}, {Ferrarese}, {Mei}, {Tonry}, \& {West}}]{Jordan2009}
{Jord{\'a}n}, A., {Peng}, E.~W., {Blakeslee}, J.~P., {et~al.} 2009, \apjs, 180, 54

\bibitem[{{Kim} \& {Lee}(2021)}]{Kim&lee2021}
{Kim}, Y.~J. \& {Lee}, M.~G. 2021, \apj, 923, 152

\bibitem[{{Kirby} {et~al.}(2013){Kirby}, {Cohen}, {Guhathakurta}, {Cheng}, {Bullock}, \& {Gallazzi}}]{Kirby2013}
{Kirby}, E.~N., {Cohen}, J.~G., {Guhathakurta}, P., {et~al.} 2013, \apj, 779, 102

\bibitem[{{Kourkchi} \& {Tully}(2017)}]{Kourkchi2017}
{Kourkchi}, E. \& {Tully}, R.~B. 2017, \apj, 843, 16

\bibitem[{{Makarov} {et~al.}(2014){Makarov}, {Prugniel}, {Terekhova}, {Courtois}, \& {Vauglin}}]{Makarov2014}
{Makarov}, D., {Prugniel}, P., {Terekhova}, N., {Courtois}, H., \& {Vauglin}, I. 2014, \aap, 570, A13

\bibitem[{{Marleau} {et~al.}(2024){Marleau}, {Duc}, {Poulain}, {M{\"u}ller}, {Lim}, {Durrell}, {Habas}, {S{\'a}nchez-Janssen}, {Paudel}, \& {Fensch}}]{Marleau2024}
{Marleau}, F.~R., {Duc}, P.-A., {Poulain}, M., {et~al.} 2024, \aap, 690, A339

\bibitem[{{Mei} {et~al.}(2007){Mei}, {Blakeslee}, {C{\^o}t{\'e}}, {Tonry}, {West}, {Ferrarese}, {Jord{\'a}n}, {Peng}, {Anthony}, \& {Merritt}}]{Mei2007}
{Mei}, S., {Blakeslee}, J.~P., {C{\^o}t{\'e}}, P., {et~al.} 2007, \apj, 655, 144

\bibitem[{{Mei} {et~al.}(2005){Mei}, {Blakeslee}, {Tonry}, {Jord{\'a}n}, {Peng}, {C{\^o}t{\'e}}, {Ferrarese}, {Merritt}, {Milosavljevi{\'c}}, \& {West}}]{Mei2005b}
{Mei}, S., {Blakeslee}, J.~P., {Tonry}, J.~L., {et~al.} 2005, \apjs, 156, 113

\bibitem[{{Miyazaki} {et~al.}(2018){Miyazaki}, {Komiyama}, {Kawanomoto}, {Doi}, {Furusawa}, {Hamana}, {Hayashi}, {Ikeda}, {Kamata}, {Karoji}, {Koike}, {Kurakami}, {Miyama}, {Morokuma}, {Nakata}, {Namikawa}, {Nakaya}, {Nariai}, {Obuchi}, {Oishi}, {Okada}, {Okura}, {Tait}, {Takata}, {Tanaka}, {Tanaka}, {Terai}, {Tomono}, {Uraguchi}, {Usuda}, {Utsumi}, {Yamada}, {Yamanoi}, {Aihara}, {Fujimori}, {Mineo}, {Miyatake}, {Oguri}, {Uchida}, {Tanaka}, {Yasuda}, {Takada}, {Murayama}, {Nishizawa}, {Sugiyama}, {Chiba}, {Futamase}, {Wang}, {Chen}, {Ho}, {Liaw}, {Chiu}, {Ho}, {Lai}, {Lee}, {Jeng}, {Iwamura}, {Armstrong}, {Bickerton}, {Bosch}, {Gunn}, {Lupton}, {Loomis}, {Price}, {Smith}, {Strauss}, {Turner}, {Suzuki}, {Miyazaki}, {Muramatsu}, {Yamamoto}, {Endo}, {Ezaki}, {Ito}, {Kawaguchi}, {Sofuku}, {Taniike}, {Akutsu}, {Dojo}, {Kasumi}, {Matsuda}, {Imoto}, {Miwa}, {Suzuki}, {Takeshi}, \& {Yokota}}]{Miyazaki2018}
{Miyazaki}, S., {Komiyama}, Y., {Kawanomoto}, S., {et~al.} 2018, \pasj, 70, S1

\bibitem[{{Moresco} {et~al.}(2022){Moresco}, {Amati}, {Amendola}, {Birrer}, {Blakeslee}, {Cantiello}, {Cimatti}, {Darling}, {Della Valle}, {Fishbach}, {Grillo}, {Hamaus}, {Holz}, {Izzo}, {Jimenez}, {Lusso}, {Meneghetti}, {Piedipalumbo}, {Pisani}, {Pourtsidou}, {Pozzetti}, {Quartin}, {Risaliti}, {Rosati}, \& {Verde}}]{Moresco2022}
{Moresco}, M., {Amati}, L., {Amendola}, L., {et~al.} 2022, Living Reviews in Relativity, 25, 6

\bibitem[{{Pietrinferni} {et~al.}(2021){Pietrinferni}, {Hidalgo}, {Cassisi}, {Salaris}, {Savino}, {Mucciarelli}, {Verma}, {Silva Aguirre}, {Aparicio}, \& {Ferguson}}]{Adriano2021}
{Pietrinferni}, A., {Hidalgo}, S., {Cassisi}, S., {et~al.} 2021, \apj, 908, 102

\bibitem[{{Poulain} {et~al.}(2021){Poulain}, {Marleau}, {Habas}, {Duc}, {S{\'a}nchez-Janssen}, {Durrell}, {Paudel}, {Ahad}, {Chougule}, {M{\"u}ller}, {Lim}, {B{\'\i}lek}, \& {Fensch}}]{Poulain2021}
{Poulain}, M., {Marleau}, F.~R., {Habas}, R., {et~al.} 2021, \mnras, 506, 5494

\bibitem[{{Raimondo}(2009)}]{Raimondo2009}
{Raimondo}, G. 2009, \apj, 700, 1247

\bibitem[{{Raimondo} {et~al.}(2005){Raimondo}, {Brocato}, {Cantiello}, \& {Capaccioli}}]{Raimondo2005}
{Raimondo}, G., {Brocato}, E., {Cantiello}, M., \& {Capaccioli}, M. 2005, \aj, 130, 2625

\bibitem[{{Riess} {et~al.}(2022){Riess}, {Yuan}, {Macri}, {Scolnic}, {Brout}, {Casertano}, {Jones}, {Murakami}, {Anand}, {Breuval}, {Brink}, {Filippenko}, {Hoffmann}, {Jha}, {D'arcy Kenworthy}, {Mackenty}, {Stahl}, \& {Zheng}}]{Reiss2022}
{Riess}, A.~G., {Yuan}, W., {Macri}, L.~M., {et~al.} 2022, \apjl, 934, L7

\bibitem[{{Rodr{\'\i}guez-Beltr{\'a}n} {et~al.}(2024){Rodr{\'\i}guez-Beltr{\'a}n}, {Cervi{\~n}o}, {Vazdekis}, \& {Beasley}}]{Rodriguez2024}
{Rodr{\'\i}guez-Beltr{\'a}n}, P., {Cervi{\~n}o}, M., {Vazdekis}, A., \& {Beasley}, M.~A. 2024, \aap, 686, A62

\bibitem[{{Schlafly} \& {Finkbeiner}(2011)}]{Schlafly2011}
{Schlafly}, E.~F. \& {Finkbeiner}, D.~P. 2011, \apj, 737, 103

\bibitem[{{Tonry} \& {Schneider}(1988)}]{Tonry1988}
{Tonry}, J. \& {Schneider}, D.~P. 1988, \aj, 96, 807

\bibitem[{{Tonry} {et~al.}(1990){Tonry}, {Ajhar}, \& {Luppino}}]{Tonry1990}
{Tonry}, J.~L., {Ajhar}, E.~A., \& {Luppino}, G.~A. 1990, \aj, 100, 1416

\bibitem[{{Tonry} {et~al.}(1997){Tonry}, {Blakeslee}, {Ajhar}, \& {Dressler}}]{Tonry1997}
{Tonry}, J.~L., {Blakeslee}, J.~P., {Ajhar}, E.~A., \& {Dressler}, A. 1997, \apj, 475, 399

\bibitem[{{Tonry} {et~al.}(2001){Tonry}, {Dressler}, {Blakeslee}, {Ajhar}, {Fletcher}, {Luppino}, {Metzger}, \& {Moore}}]{Tonry2001}
{Tonry}, J.~L., {Dressler}, A., {Blakeslee}, J.~P., {et~al.} 2001, \apj, 546, 681

\bibitem[{{Vazdekis} {et~al.}(2020){Vazdekis}, {Cervi{\~n}o}, {Montes}, {Mart{\'\i}n-Navarro}, \& {Beasley}}]{Vazdekis2020}
{Vazdekis}, A., {Cervi{\~n}o}, M., {Montes}, M., {Mart{\'\i}n-Navarro}, I., \& {Beasley}, M.~A. 2020, \mnras, 493, 5131

\bibitem[{{Villegas} {et~al.}(2010){Villegas}, {Jord{\'a}n}, {Peng}, {Blakeslee}, {C{\^o}t{\'e}}, {Ferrarese}, {Kissler-Patig}, {Mei}, {Infante}, {Tonry}, \& {West}}]{Villegas2010}
{Villegas}, D., {Jord{\'a}n}, A., {Peng}, E.~W., {et~al.} 2010, \apj, 717, 603

\bibitem[{Virtanen {et~al.}(2020)Virtanen, Gommers, Oliphant, Haberland, Reddy, Cournapeau, Burovski, Peterson, Weckesser, Bright, {van der Walt}, Brett, Wilson, Millman, Mayorov, Nelson, Jones, Kern, Larson, Carey, Polat, Feng, Moore, {VanderPlas}, Laxalde, Perktold, Cimrman, Henriksen, Quintero, Harris, Archibald, Ribeiro, Pedregosa, {van Mulbregt}, \& {SciPy 1.0 Contributors}}]{2020SciPy-NMeth}
Virtanen, P., Gommers, R., Oliphant, T.~E., {et~al.} 2020, Nature Methods, 17, 261

\bibitem[{{Weisz} {et~al.}(2014){Weisz}, {Dolphin}, {Skillman}, {Holtzman}, {Gilbert}, {Dalcanton}, \& {Williams}}]{Weisz2014}
{Weisz}, D.~R., {Dolphin}, A.~E., {Skillman}, E.~D., {et~al.} 2014, \apj, 789, 147

\bibitem[{{Worthey}(1994)}]{Worthey1994}
{Worthey}, G. 1994, \apjs, 95, 107

\end{thebibliography}

\begin{appendix}

\onecolumn
\section{\label{appendix:tables}Data and SBF measurements tables}
Here we report the properties of the HSC-SSP galaxies (Table \ref{table:1}) and the results of the SBF
analysis (Table \ref{table:2}).

\begin{table*}[htb]
\caption{List of target HSC galaxies.   }  
\label{table:1}
\setlength{\tabcolsep}{11pt}
\begin{tabular}{l l l l l l l r}
\hline\hline
  & & & & & & &\\[-9pt]
  \multicolumn{1}{c}{Name} &
  \multicolumn{1}{c}{MATLAS ID.} &
  \multicolumn{1}{c}{RA (J2000)} &
  \multicolumn{1}{c}{DEC (J2000)} &
  \multicolumn{1}{c}{$B_\mathrm{T}$} &
  \multicolumn{1}{c}{$v_{\mathrm{CMB}}$} &
  \multicolumn{1}{c}{T} & 
  \multicolumn{1}{c}{Group ID.}\\
  \multicolumn{1}{c}{} &
  \multicolumn{1}{c}{} &
  \multicolumn{1}{c}{(deg)} &
  \multicolumn{1}{c}{(deg)} &
  \multicolumn{1}{c}{(mag)} &
  \multicolumn{1}{c}{(kms$^{-1}$)} &
  \multicolumn{1}{c}{} &
  \multicolumn{1}{c}{}\\
  \multicolumn{1}{c}{(1)} &
  \multicolumn{1}{c}{(2)} &
  \multicolumn{1}{c}{(3)} &
  \multicolumn{1}{c}{(4)} &
  \multicolumn{1}{c}{(5)} &
  \multicolumn{1}{c}{(6)} &
  \multicolumn{1}{c}{(7)} &
  \multicolumn{1}{c}{(8)}\\
  [0.5ex] 
\hline
& & & & & & & \\[-9pt]
    NGC\,5813\footnotemark[1] & - & 225.296840 & 1.701980 & 11.45 & 2154 & -4.9 & 53932\footnotemark[2]\\
    PGC\,053521 & - & 224.702969 & 2.023514 & 14.64 & 2005 & -4.0 & 53932\\
    PGC\,053587 & - & 225.069056 & 2.300727 & 15.24 & 2016 & -2.0 & 53932\\
    PGC\,053636 & - & 225.262961 & 0.707490 & 15.51 & 1923 & - & 53932\\
    PGC\,1165764 & - & 225.982034 & 0.432116 & 16.81 & 1781 & - & 53932\\
    PGC\,1192611 & MATLAS\,1962 & 225.617203 & 1.364335 & 17.93 & 1727 & -& 53932\\
    PGC\,1193898 & MATLAS\,1942 & 225.219140 & 1.404918 & 16.69 & 2083 & -& 53932\\
    PGC\,1205406 & MATLAS\,1950 & 225.316184 & 1.773550 & 17.73 & 1539 & - & 53932\\
    PGC\,1208589 & MATLAS\,1955 & 225.410746 & 1.870024 & 17.63 & 2349 & -& 53932 \\
    PGC\,1230503 & MATLAS\,1979 & 225.934513 & 2.552337 & 17.31 & 1975 & - & 53932\\
    PGC\,184851 & - & 224.588164 & 1.845407 & 15.77 & 2070 & - & 53932\\
    PGC\,3350761 & MATLAS\,1945 & 225.247347 & 1.649311 & 18.34 & 2556 & -& 53932 \\
    PGC\,3350757 & MATLAS\,1954 & 225.409960 & 1.722182 & 17.83 & 2487 & - & 53932\\
    PGC\,3350778 & MATLAS\,1938 & 225.137620 & 2.230316 & 16.99 & 1510 & - & 53932\\
    PGC\,4018458 & MATLAS\,1944 & 225.247330 & 1.876714 & 18.76 & 2378 & - & 53932\\
    PGC\,4018466 & MATLAS\,1963 & 225.637999 & 1.935573 & 17.93 & 1842 & - & 53932\\
    PGC\,4074773 & - & 224.619334 & 1.542849 & 17.40 & 1696 & - & 53932\\ 
    
    NGC5839\footnotemark[1] &- & 226.364527 & 1.634810 & 13.46 & 1412 & -2.0& 53932 \\
    PGC\,1199471 & MATLAS\,2027 & 226.382612 & 1.587682 & 17.51 & 1111  & - & - \\
    PGC\,4009173 & MATLAS\,2005 & 226.202155 & 1.980828 & 17.68 & 2152 & - & 53932\\
    
    NGC5831\footnotemark[1] &-& 226.029183 & 1.219933 & 12.45 & 1824 & -4.8 & 53932\\
    PGC\,1190315 & MATLAS\,2002 & 226.178708 & 1.290919 &  16.28 & 2160   & -& 53932 \\
    PGC\,1216386 & MATLAS\,1994 & 226.102938 & 2.114612 & 16.93 & 1895 & -& 53932 \\
    \hline
    & & & & & & & \\[-9pt]
    
    NGC4753\footnotemark[1] &-& 193.091951 & -1.199610 & 10.85 & 1497 & -1.3 & 43671\footnotemark[3]\\
    PGC\,1150193 &- & 193.141851 & -0.167855 & 18.09 & 1413 & - & 43671\\
    PGC\,135809 &- & 192.519806 & -0.232687 & 17.59 & 1089 & - & 43671\\
    PGC\,183384 & MATLAS\,1522 & 193.140620 & -1.730207 & 15.76 & 1464 & - & 43671\\
    PGC\,3295853 & MATLAS\,1501 &192.113308 & -1.667457 & 19.07 & 1566  & -& 43671 \\
    PGC\,44066 &- & 193.914018 & -0.263410 & 14.75 & 1441 & - & 43671\\
    
    NGC4690 & -&191.981333 & -1.656050 & 13.87 & 3103 & -3 & 43202\\
    PGC\,1115603 & MATLAS\,1491 & 191.801240 & -1.569420 & 16.73 & 3167  & -& 43202 \\
    PGC\,1124896 & MATLAS\,1477 & 191.558178 & -1.178323 & 17.05 & 3067   & - & -\\
    \hline
    & & & & & & & \\[-9pt]
    
    IC0745\footnotemark[1] &- & 178.551187 & 0.136719 & 14.17 & 1143 & -2.2& 37339\footnotemark[4] \\
    GAMA559879 &- & 178.551042 & -0.481856 & 17.20 & 2398   & - & -\\
    GAMA583929 &- & 177.378693 & -0.025551 & 19.09 & 2762   & -& -\\
    IC\,0753 &- & 179.803649 & -0.523923 & 14.48 & 6768   & -1.8 & -\\
\hline
\end{tabular}
\footnotetext[1]{}{ $^1$Reference sample of galaxies from \citetalias{Tonry2001}.}
\footnotetext[2]{}{ $^2$Virgo III group.}
\footnotetext[3]{}{ $^3$Virgo II group.}
\footnotetext[4]{}{ $^4$Leo II group.}
\footnotetext[5]{}{The horizontal lines delimit the three regions of the sky, shown in Fig. \ref{fig:spatial_distr}. For each galaxy, we list the galaxy name (Col. 1); the galaxy name from the MATLAS survey if available (Col. 2); celestial coordinates in degrees (Col. 3 and 4); total $B$-band magnitude from Hyperleda\footnote{http://atlas.obs-hp.fr/hyperleda/} \citep{Makarov2014} not corrected for extinction (Col. 5); velocity $v_{\mathrm{CMB}}$ from the NASA Extragalactic Database (NED) in the Cosmic Microwave Background (CMB) rest-frame (Col. 6); morphological $T_{\mathrm{type}}$ (Col. 7) from Hyperleda, only for galaxies brighter than $B_\mathrm{T}\sim15$ mag; group identifier (Col. 8), according to \citet{Kourkchi2017}.}
\end{table*}

\begin{table*}
\caption{SBF, colors and distances for the HSC-SSP galaxies in our sample.}  
\label{table:2}
\setlength{\tabcolsep}{11pt}
\begin{tabular}{l l l l l l l r}
\hline\hline
\\
  \multicolumn{1}{c}{Name} &
  \multicolumn{1}{c}{($g-i$)}&
  \multicolumn{1}{c}{$\overline{m_\mathrm{i}}$}&
  \multicolumn{1}{c}{$\overline{m}-\overline{M}$}&
  \multicolumn{1}{c}{$d$}&
  \multicolumn{1}{c}{$r_{\mathrm{in}}$/$r_{\mathrm{out}}$}&
  \multicolumn{1}{c}{$fraction_{\mathrm{mask}}$}&
  \multicolumn{1}{c}{Flag}\\
  \multicolumn{1}{c}{} &
  \multicolumn{1}{c}{(mag)}&
  \multicolumn{1}{c}{(mag)}&
  \multicolumn{1}{c}{(mag)}&
  \multicolumn{1}{c}{(Mpc)}&
  \multicolumn{1}{c}{(arcsec)}&
  \multicolumn{1}{c}{}&
  \multicolumn{1}{c}{}\\
  \multicolumn{1}{c}{(1)} &
  \multicolumn{1}{c}{(2)}&
  \multicolumn{1}{c}{(3)}&
  \multicolumn{1}{c}{(4)}&
  \multicolumn{1}{c}{(5)}&
  \multicolumn{1}{c}{(6)}&
  \multicolumn{1}{c}{(7)}&
  \multicolumn{1}{c}{(8)}\\
  [0.5ex] 
\hline\\
  NGC\,5813 &  1.206 (0.004) & 31.58 (0.06) & 31.92 (0.12) & 24.2 (1.3) & 7/68 & 0.46 & q1 \\
  PGC\,053521 &  0.991 (0.007) & 30.92 (0.08) & 32.02 (0.12) & 25.4 (1.4) & 2/17 & 0.30 & q2 \\
  PGC\,053587 &  0.998 (0.013) & 30.98 (0.28) & 32.07 (0.29) & 26.0 (3.5) & 2/17 & 0.30 & q2 \\
  PGC\,053636 &  0.921 (0.017) & 30.74 (0.10) & 32.01 (0.13) & 25.2 (1.6) & 2/17 & 0.30 & q2 \\
  PGC\,1165764 &  0.455 (0.022) & 29.35 (0.09) & 32.04 (0.13) & 25.6 (1.5) & 2/17 & 0.30 & q2 \\
  PGC\,1192611 &  0.798 (0.026) & 30.73 (0.22) & 32.23 (0.24) & 28.0 (3.2) & 0.17/6 & 0.10 & q3\\
  PGC\,1193898 &  0.988 (0.019) & 30.95 (0.09) & 32.08 (0.13) & 26.0 (1.5) & 0.17/23.7 & 0.50 & q2\\
  PGC\,1205406 &  0.897 (0.039) & 30.57 (0.18) & 31.90 (0.20) & 23.9 (2.3) & 0.17/12 & 0.32 & q3\\  
  PGC\,1208589 &  0.887 (0.028) & 30.45 (0.17) & 31.79 (0.20) & 22.8 (2.1) & 0.17/7.4 & 0.32 & q2\\
  PGC\,1230503 &  0.920 (0.016) & 31.08 (0.11) & 32.36 (0.14) & 29.6 (1.9) & 2/17 & 0.30 & q2 \\
  PGC\,184851 &  1.050 (0.007) & 31.09 (0.07) & 32.02 (0.12) & 25.4 (1.4) & 2/17 & 0.30 & q2 \\
  PGC\,3350761 &  1.105 (0.033) & 31.68 (0.21) & 32.46 (0.23) & 31.1 (3.4) & 0.17/6.7 & 0.34 & q2\\
  PGC\,3350757 &  1.047 (0.023) & 31.18 (0.14) & 32.16 (0.17) & 27.0 (2.1) & 2/17 & 0.30 & q2 \\
  PGC\,3350778 &  0.932 (0.018) & 30.72 (0.06) & 31.98 (0.11) & 24.9 (1.2) & 2/17 & 0.30 & q2 \\
  PGC\,4018458 &  0.872 (0.037) & 31.05 (0.14) & 32.42 (0.17) & 30.5 (2.4) & 0.17/6 & 0.18 & q3\\
  PGC\,4018466 &  0.936 (0.043) & 30.96 (0.23) & 32.20 (0.25) & 27.6 (3.18) & 0.17/6.7 & 0.48 & q2\\
  PGC\,4074773 &  0.858 (0.017) & 31.04 (0.16) & 32.25 (0.18) & 28.1 (2.4) & 2/17 & 0.30 & q2 \\
  NGC\,5839 & 1.131 (0.004) & 31.50 (0.17) & 32.17 (0.19) & 27.2 (2.4) & 17/34 & 0.35 & q1\\
  PGC\,1199471 &  0.927 (0.023) & 30.75 (0.08) & 32.01 (0.12) & 25.2 (1.4) & 0.17/6.7 & 0.15 & q2\\
  PGC\,4009173 & 0.688 (0.028) & 29.82 (0.16) & 31.55 (0.18) & 20.4 (1.7) & 0.17/6.7 & 0.23 & q3\\
  NGC\,5831 & 1.133 (0.008) & 31.48 (0.09) & 32.17 (0.13) & 27.1 (1.6) & 20/56 & 0.44 & q1\\
  PGC\,1190315 &  1.043 (0.07) & 30.79 (0.09) & 31.78 (0.14) & 22.7 (1.5) & 0.17/8.4 & 0.07 & q3\\
  PGC\,1216386 &  0.918 (0.017) & 30.89 (0.08) & 32.17 (0.12) & 27.1 (1.5) & 0.17/8.4 & 0.26 & q2\\
  \textbf{Group 53932}\footnotemark[1] & - & - &32.09 (0.12)& 26.2 (1.5) & - &- &- \\
  \hline
    & & & & & & & \\[-9pt]
    
  NGC\,4753 &  0.958 (0.002) & 30.77 (0.10) & 31.92 (0.13) & 24.3 (1.5) & 7/17 & 0.53 & q1\\
  PGC\,1150193 &  0.363 (0.049) & 28.97 (0.14) & 32.01 (0.17) & 25.2 (2.1) & 0.3/6 & 0.38 & q2\\
  PGC\,135809 &  0.667 (0.053) & 29.63 (0.18) & 31.37 (0.20) & 18.8 (1.8) & 0.3/6 & 0.38 & q2\\
  PGC\,183384 & 0.780 (0.013) & 30.0 (0.10) & 31.51 (0.13) & 20.1 (1.2) & 0.3/56 & 0.43 & q3\\
  PGC\,3295853 & 0.750 (0.041) & 30.20 (0.23) & 31.77 (0.25) & 22.5 (2.6) & 0.17/5 & 0.15 & q3\\
  PGC\,44066 & 0.5012 (0.013) & 29.46 (0.19) & 31.72 (0.21) & 22.1 (2.1) & 0.3/56 & 0.43 & q3\\
  \textbf{Group 43671}\footnotemark[1] & - & - &31.77 (0.17)& 22.6 (1.9) & - &- &- \\
  NGC4690 & 0.975 (0.003) & 31.515 (0.06) & 32.63 (0.11) & 33.4 (1.7) & 14/22 & 0.15 & q1\\
  PGC\,1115603 & 0.736 (0.026) & 30.67 (0.26) & 32.26 (0.27) & 28.4 (3.6) & 2.5/6.7 & 0.60 & q3\\
  PGC\,1124896 & 0.570 (0.025) & 30.22 (0.34) & 32.22 (0.35) & 27.8 (4.5) & 0.17/6.7 & 0.35 & q3\\
  \hline
    & & & & & & & \\[-9pt]
    
  IC\,0745 & 0.828 (0.008) & 30.21 (0.06) & 31.62 (0.11) & 21.1 (1.1) & 13/27 & 0.44 & q1\\
  GAMA559879 & 0.546 (0.037) & 30.01 (0.2) & 32.09 (0.22) & 26.2 (2.7) & 0.3/6.7 & 0.39 &q3\\
  GAMA583929 & 0.586 (0.047) & 30.78 (0.64) & 32.73 (0.64) & 35.1 (10.4) & 0.17/5 & 0.28 & q3\\
  \textbf{Group 37339}\footnotemark[1] & - & - &31.62 (0.11)& 21.1 (1.1) & - &- &- \\
\hline
\end{tabular}
\footnotetext[1]{}{$^1$ Group ID From \citet{Kourkchi2017}.}

\footnotetext[1]{}{We report the galaxy name (Col. 1); the ($g-i$)$_{\mathrm{HSC}}$ color and its uncertainty (Col. 2). Colors and SBF magnitudes in the table are uncorrected for Galactic extinction; the SBF magnitude (Col. 3); the distance modulus and the distance in Mpc (Col. 4 and 5); the internal and external radii of the annulus adopted for the SBF measurement (Col. 6); the fraction of masked pixels in the annulus (Col. 7); the quality flag (Col. 8, $q1\equiv$ excellent quality, $q2\equiv$  good quality, and $q3\equiv$ poor quality.) The horizontal lines delimit the three fields, as seen in Fig.\,\ref{fig:spatial_distr}. The $1\sigma$ uncertainties on the measurements are in parenthesis..}
\end{table*}

\end{appendix}

\end{document}